\documentstyle[prd,aps,epsfig]{revtex}

\begin{document} 
 
\title{%
\vskip-6pt \hfill {\rm\normalsize UCLA/00/TEP/5} \\ 
\vskip-6pt \hfill {\rm\normalsize MPI-PhT/2000-05} \\ 
\vskip-6pt \hfill {\rm\normalsize March 2000} \\
\vskip-9pt~\\
Measurement of the gluon PDF at small $x$ with neutrino telescopes
}
 
\author{Graciela Gelmini\rlap{,}{$^{1}$} Paolo Gondolo\rlap{,}{$^{2}$} 
Gabriele Varieschi$^{1,3}$} 
 
\address{
~\\
${}^{1}$ Dept. of Physics and Astronomy, UCLA (University of 
California, Los Angeles)\\
405 Hilgard Ave., Los Angeles CA 90095, USA\\
{\rm gelmini,variesch@physics.ucla.edu}
\\~\\
${}^{2}$ Max-Planck-Institut f\"ur Physik (Werner-Heisenberg-Institut) 
\\
F\"ohringer Ring 6, 80805 M\"unchen, Germany\\
{\rm gondolo@mppmu.mpg.de}
\\~\\
${}^{3}$ Dept. of Physics, Loyola Marymount University\\
7900 Loyola Blvd., Los Angeles CA 90045, USA\\
{\rm gvariesc@lmumail.lmu.edu}
\\~\\
} 
 
\maketitle
 
\begin{abstract} 
  We analyze the possibility that neutrino telescopes may provide
  an experimental determination of the slope $\lambda$ of the gluon
  distribution in the proton at momentum fractions $x$ smaller than the
  accelerator reach. The method is based on a linear relation between $\lambda$
  and the spectral index (slope) of the down-going atmospheric muon flux above
  100 TeV, for which there is no background. Considering the uncertainties in 
the charm production cross section and in the cosmic ray composition, we
  estimate the error on the measurement of $\lambda$
   through this method, excluding the experimental error of the telescopes, to 
be $\pm~0.2$.
\end{abstract}

 
\section{Introduction} 
 
\label{sect:intro4} 
 
Atmospheric neutrinos and muons are the most important source of background for
present and future neutrino telescopes, which are expected to open a new
window in astronomy by detecting neutrinos from astrophysical sources.
 
At energies above 1 TeV, atmospheric lepton fluxes have a prompt component 
consisting of neutrinos and
muons created in semileptonic decays of charmed particles, as opposed to the
conventional leptons coming from decays of pions and kaons. Thus a model
for charm production and decays in the atmosphere is required.

We base our model on QCD, the theoretically preferred model, to compute the 
charm production. We use a next-to-leading order perturbative QCD (NLO pQCD) 
calculation of charm
production in the atmosphere, followed by a full simulation of particle
cascades generated with PYTHIA routines \cite{PYTHIA}.
 
We have already examined the prompt muon and neutrino fluxes in two
previous papers \cite{GGV1,GGV2} (called GGV1 and GGV2 from now on).

In our first paper \cite{GGV1}, we found that the NLO pQCD approach produces 
fluxes in the bulk of older predictions (not
based on pQCD) as well as of the recent pQCD semianalytical analysis of
Pasquali, Reno and Sarcevic \cite{PRS}. We also explained the reason of
the low fluxes of the TIG model \cite{TIG}, the first to use pQCD in this
context, which were due to the chosen extrapolation of the gluon partonic
distribution function (PDF) at small momentum fractions $x$, and we confirmed 
the overall validity of their pioneering approach to the problem.
 
In our second paper \cite{GGV2}, we analyzed in detail the dependence of
the fluxes on the extrapolation of the gluon PDF at small $x$, which, according
to theoretical models, is assumed to be a power law with exponent $\lambda$,
\begin{equation} 
x g(x) \sim x^{-\lambda},  
\label{eq:4.1} 
\end{equation} 
with $\lambda$ in the range 0--0.5. Particle physics experiments are yet unable
to determine the value of $\lambda$ at $x<10^{-5}$. We found that the choice of
different values of $\lambda$ at $x<10^{-5}$ leads to a wide range of final
background fluxes at energies above 10$^5$ GeV.
 
Due to this result, in GGV2 we suggested the possibility of measuring $\lambda$ 
through the
atmospheric leptonic fluxes at energies above $10^5$ GeV, not the absolute 
fluxes, because of their large
theoretical error, but rather through their spectral index (i.e.\ the 
``slope''). In particular, we now propose to use the slope of the flux of 
down-going muons. 

We want to stress that we are proposing to use {\it down-going muons}, at 
energies $E_{\mu} \gtrsim 100$ TeV, where prompt muons dominate over 
conventional ones, and not up-going neutrino-induced muons whose flux
is orders of magnitude smaller. 
While an important contribution to up-going muons is expected from astrophysical 
neutrinos, there is no background for down-going atmospheric muons.
 
In this paper we further investigate the possibility of measuring $\lambda$, in 
the more general context of an overall error
analysis of our model.
 
We can identify five potential causes of uncertainty in our final results. The 
first
one is the possible presence of large logarithms of the type $\alpha_s \ln
p^2_T$ and $\alpha_s \ln s$ (the latter are the so called ``$\ln(1/x)$''
terms). 
The second is the treatment of the multiplicity in the production of $c\bar{c}$ 
at high energies.

The third one consists of all the sources of uncertainty hidden in the
treatment of particle cascades generated by PYTHIA. 
The fourth one is the uncertainty in the NLO pQCD charm production model we use. 
This includes the dependence of the fluxes on the three parameters of the model 
and the PDF's used.
The fifth and final one is
the choice of the primary cosmic ray flux, which is the input of our
simulation.
Of all these potential sources of uncertainty we conclude that only the last two 
are relevant.

We deal with these five potential sources of error in turn.
In Sect.~\ref{sect:logs4} we address the question of the large logarithms
$\alpha_s \ln p^2_T$ and $\alpha_s \ln s$, and in Sect.~\ref{sect:mult4} we
analyze the problem of multiplicity in our charm production mechanism.

In Sect.~\ref{sect:charm4} we consider the uncertainties due to the cascade 
generation by PYTHIA and to our NLO pQCD charm
production (the core of our analysis). Here we evaluate the errors due to the 
parameters of the
model, errors that affect the charm production cross section, the final
differential and integrated fluxes and their spectral indices. We also
determine how the final results (fluxes and their spectral indices) are affected 
by
the choice of different extrapolations of the PDF's at $x<10^{-5}$.
 
We are finally ready in Sect.~\ref{sect:lambda4} to discuss how $\lambda$ could 
be measured. We study the dependence on the different
extrapolations of $\lambda$ at $x<10^{-5}$, we consider the
spectral indices and, using the discussion of Sect.~\ref{sect:charm4}, we
provide an estimate of the errors on these indices and examine the feasibility
of an experimental determination of $\lambda$ at $x<10^{-5}$ with neutrino
telescopes.
 
Finally in Sect.~\ref{sect:composition4} we discuss the error on the
determination of $\lambda$ coming from the uncertainties in the elemental
composition of the cosmic ray flux.


The determination of $\lambda$ at small $x<10^{-5}$ is important because in this 
range
saturation, unitarity and shadowing effects should become important.
The PDF sets we use have been derived without including saturation effects.
Even if this procedure seems to work very well in the HERA regime (where there
might be some indications of saturation already, see e.g. \cite{Mueller}), here 
we are
extrapolating the resulting gluon PDF's  at even smaller $x$ values where 
saturation
may become important. With respect to unitarity, using the expression of the 
Froissart bound on the gluon structure functions given in Eq.~31 of 
Ref.~\cite{Ayala1} we see that the extrapolated gluon PDF's we use, with 
$\lambda=0.4-0.5$, violate this bound at $x$ values between 0.5 $\times 10^{-7}$ 
and 1 $\times 10^{-7}$, 
for the characteristic momenta $Q^2 \simeq m_c^2 \simeq 3$ GeV$^2$ we have, 
which corresponds to leptonic energies of $1 - 2 \times 10^{6}$ GeV. Always 
using the Froissart bound
on the gluon PDF as given in the Eq.~31 of Ref.~\cite{Ayala1}, 
the gluon PDF`s extrapolated with $\lambda \leq 0.3$ encounter this bound 
 at $x < 10^{-8}$, what corresponds to leptonic energies larger than $10^8$ GeV, 
beyond the energy range 
relevant in this paper. Shadowing of the gluons 
in the atmospheric nucleons and nuclei, which we have not included here,  could 
decrease the amplitude of the gluon PDF's by about $10\%$ at $x \simeq 10^{-5}$ 
and up to 
as much as $30$ to $40\%$ at  $x \simeq 10^{-7}$ (see e.g. \cite{Ayala2}), what 
would also change the  effective value of $\lambda$. There are no shadowing 
effects in the cosmic ray nucleons and nuclei. Only the dominant $x$ of the 
gluons in the atmosphere is small 
in our calculation, while the dominant $x$ of the partons 
in the cosmic rays is large. Thus shadowing effects
do not depend on the unknown composition of cosmic rays, 
but could only be  important for the atmospheric partons. The reason for the 
different characteristic values of $x$ in the target and projectile
partons is the following (for more details see GGV2). 
Due to the steep decrease with
increasing energy of the incoming flux of cosmic rays, only the most energetic
charm quarks produced count and those come from the interaction of
projectile partons carrying a large fraction of the incoming nucleon momentum.
Thus, the characteristic $x$ of the projectile parton,  $x_1$, is
large. It is $x_1 \simeq O(10^{-1})$.  We can, then, inmediately understand
that very small parton momentum fractions  are
needed in our calculation, because typical partonic center of mass energies
$\sqrt{\hat s}$ are close to the $c \bar c$ threshold, $2m_c \simeq 2$ GeV,
(since the differential $c \bar c$ production cross section decreases with 
increasing ${\hat s}$)
while the total center of mass energy squared is $s = 2 m_N E$ (with $m_N$ the
nucleon mass, $m_N \simeq 1$ GeV). Calling $x_2$ the momentum fraction 
of the target parton (in a nucleous of the atmosphere),
then, $x_1 x_2 \equiv \hat s/s = 4 m_c^2/(2 m_N E) \simeq$ GeV/$E$. Thus,
$x_2 \simeq O$(GeV/0.1 $E$), where $E$ is the energy per nucleon of the
incoming cosmic ray in the lab. frame. The characteristic energy $E_c$ of the
charm quark and  the dominant leptonic energy $E_l$ in the fluxes  are
$E_l \simeq E_c \simeq 0.1 E$, thus $x_2 \simeq O$(GeV/ $E_l$).

\section{Importance of the $\alpha_s \ln(1/x)$ terms}
\label{sect:logs4}

We address here a concern that has been expressed to us several times, about
the applicability of perturbative QCD calculations, mostly done for accelerator
physics, to the different kinematic domain of cosmic rays. 

Contrary to the case in accelerators, we do not have the uncertainty present in
the differential cross sections \cite{MNR} when the transverse momentum $p_T$
is much larger than $m_c$, uncertainty which is due to the presence of
large logarithms of $(p^2_T + m_c^2)/m_c^2$. The reason is that we do not have
a forward cut in acceptance, and so the characteristic transverse charm
momentum in our simulations is of the order of the charm mass, $p_T \simeq
O(m_c)$, and not $p_T \gg O(m_c)$ as in accelerator experiments.

We may however, depending on the steepness of the gluon structure function
$\lambda$, have large logarithms of the type $\alpha_s \ln s$, known as
``$\ln(1/x)$'' terms (here $x\simeq \sqrt{4m_{c}^{2}/s}$ is the average value
of the hadron energy fraction needed to produce the $c\bar{ c}$ pair at
hadronic center of mass energy squared $s$). These ``$\ln(1/x)$'' terms arise
when the t-channel gluon exchange becomes large, and eventually they have to be 
resummed.
Although techniques exist for resumming these logarithms~\cite{resumlog}, we
have not done it. On the other hand we have phenomenologically altered the
behavior of the parton distribution functions at small $x$ by imposing a power
law dependence of the form $ x f(x) \sim x^{-\lambda}$. This is analogous to
resumming the $\ln(1/x)$ terms
 in a universal fashion and absorbing them in an improved
evolution equation for the gluon density (such as the 
Balitskyi\u{i}-Fadin-Kuraev-Lipatov (BFKL) evolution
equation)~\cite{mangano97}, a procedure which increases $\lambda$.
  For sufficiently large $\lambda$, the large $\ln(1/x)$ terms should not be 
present.  

To find if our NLO $c\bar{c}$ cross sections are dominated by the $\ln(1/x)$
terms, we have used the following qualitative criterion \cite{crit}.  We have
plotted the ratio
\begin{equation}
R = \frac{ \sigma_{NLO} - \sigma_{LO} }
         { \sigma_{LO} \alpha_s \ln(s/m_c^2)/\pi } 
\label{eq:4.2}
\end{equation}
as a function of the beam energy $E$. If the ratio is constant we are dominated 
by the $\ln(1/x)$ terms
and if it decreases we are not. The
good behavior is a decreasing $R$. Figure~1 shows indeed that up to highest
energy we consider in this paper, i.e.\ $10^{11}$ GeV, $R$ decreases for
$\lambda \gtrsim 0.2$, but is roughly constant for smaller $\lambda$'s. This
indicates that we are not dominated by the $\ln(1/x)$ terms provided $\lambda
\gtrsim 0.2$.

Clearly, this test involving the $R$ ratio does not say anything about 
$ln(1/x)$ higher order corrections. One can only argue that if the $ln(1/x)$ 
terms
are not dominant at NLO (for $R$ 
decreasing with energy) the corresponding $[ln(1/x)]^n$ terms may also be 
non-dominant
in higher order corrections. In any event, the data on charm 
production that 
could be inferred at $x <10^{-5}$, from the slope of atmospheric muon fluxes,
really give information on the product of the gluon PDF and the parton cross 
section and a measurement of this product is useful. One can expect that 
the $ln(1/x)$ terms at higher order may be better understood by the time the 
data will come.

\section{Multiplicity in charm production}
\label{sect:mult4} 

Another concern is the fact that at high energies the charm production cross
section we use is sometimes larger than the total
$pp$ cross section. At first sight this seems absurd, but we show here that
 the cross section we use is the inclusive cross section, which contains the 
charm multiplicity, i.e.\ it counts the
number of $c\overline{c}$ pairs produced, and so can be larger than the total
cross section. On the other hand, the contribution of $c\overline{c}$ producing 
events to the  total $pp$ cross section,
i.e.\ the cross section for producing at least one $c\overline{c}$ pair, is
always smaller than the total $pp$ cross section.

We call $\sigma_{QCD}$ the perturbative QCD cross section of $c\bar{c}$ pair
production in $pp$ collisions,
\begin{equation}
  \sigma_{QCD} = \sum_{ij} \sigma_{QCD}(ij\to c\bar{c}) ,
\label{eq:4.3}
\end{equation}
where the sum is over the partons $i$ and $j$ in the colliding nucleons, and
\begin{equation}
    \sigma_{QCD}(ij\to c\bar{c}) = \int dx_1 dx_2 dQ^2 
\frac{d\hat{\sigma}(ij\to c\bar{c})}{dQ^2} f_i(x_1,\mu_F^2)
f_j(x_2,\mu_F^2) .
\label{eq:4.4}
\end{equation}
Here $d\hat{\sigma}(ij\to c\bar{c})/dQ^2$ is the $ij\to c\bar{c}$ parton
scattering cross section, $Q^2$ is the four-momentum transfer squared, $x_i$ is
the fraction of the momentum of the parent nucleon carried by parton $i$, and
$f_i(x,\mu_F^2)$ is the usual parton distribution function for parton momentum
fraction $x$ and factorization scale $\mu_F$.

In the scattering of each pair of partons (one parton from the target and
one from the projectile) only one $c\bar{c}$ pair may be produced, but the 
number
of parton pairs interacting in each nucleon-nucleon collision is in general not
limited to one and it increases with the number of partons $f(x,\mu_F^2) dx$ in
each nucleon.

For $\lambda$ close to 0.5, $\sigma_{QCD}$ becomes larger than the total $pp$
cross section $\sigma_{pp} \sim 200~{\rm mb}$ at $E_p \sim 10^{10}$ GeV.  It
is obvious therefore that our results at high energy and large $\lambda$ are
unphysical, unless multiplicity is taken into account. In fact, multiparton
interactions should be taken into account already at a smaller cross section of
order 10 mb, as determined in studies of double parton scattering~\cite{DP}.

In order to incorporate multiparton scatterings into our analysis, we use an
impact-parameter representation for the scattering amplitude, and ignore
spin-dependent effects (cfr.~\cite{dp87}).  Assuming the validity of
factorization theorems, the mean number of parton-parton interactions $ij\to
c\bar{c}$ in the collision of two protons at impact parameter $\vec{b}$ 
is given by
\begin{equation}
n_{c\bar{c}}(\vec{b}) = \sum_{ij} \int d^2b' dx_1 dx_2 dQ^2 
\frac{d\hat{\sigma}(ij\to c\bar{c})}{dQ^2} f_i(x_1,\mu_F^2,\vec{b'})
f_j(x_2,\mu_F^2,\vec{b}+\vec{b'}) ,
\label{eq:4.5}
\end{equation}
where $f_i(x,\mu_F^2,\vec{b}) d^2b$ is the number of partons $i$ in the interval
($x$, $x+dx$) and in the transverse area element $d^2b$ at a distance 
$\vec{b}$ from the center of the proton. For simplicity of notation
 we drop the vector symbol in $\vec{b}$ and write $b$ from now on.

If $n_{c\bar{c}}(b) \ll 1$, $n_{c\bar{c}}(b)$ is the probability of producing a
$c\bar{c}$ pair in a $pp$ collision at impact parameter $b$.  If
$n_{c\bar{c}}(b) \ge 1$, $n_{c\bar{c}}(b)$ is just the mean value of $k$, the
number of $c\bar{c}$ pairs produced, at impact parameter $b$.  Let the
probability of $k$ scatterings $ij\to c\bar{c}$ in a $pp$ collision at impact
parameter $b$ be $P_{kc\bar{c}}(b)$.  Then
\begin{equation}
n_{c\bar{c}}(b) = \sum_{k=0}^\infty k P_{kc\bar{c}}(b) .
\label{eq:4.6}
\end{equation}

The $k$-tuple parton cross section is obtained by integrating the
probability of exactly $k$ parton scatterings $P_{kc\bar{c}}(b)$ over the
impact parameter $b$,
\begin{equation}
  \sigma_{kc\bar{c}} = \int d^2b ~P_{kc\bar{c}}(b),
\label{eq:4.8}
\end{equation}
 the inclusive cross section for charm production is, thus, $\sigma_{c\bar{c}\rm 
incl} = \sum_k  k \sigma_{kc\bar{c}}$ and the contribution of charm producing 
processes to the
total cross section is $\sigma_{c\bar{c}} = \sum_k  \sigma_{kc\bar{c}}$
for $k \not= 0$.

In our evaluation of charm production by cosmic ray
interactions in the atmosphere, we must count the number of $c\bar{c}$ pairs
produced in the $pp$ collision. So we are precisely interested in the inclusive 
cross section $\sigma_{c\bar{c}\rm incl}$, which includes the
number $k$ of $c\bar{c}$ pairs produced per collision (the multiplicity).  We
find
\begin{equation}
  \sigma_{c\bar{c}\rm incl} = \sum_k k \sigma_{kc\bar{c}} = \int d^2b \sum_k k
  P_{kc\bar{c}}(b)  = \int  d^2b ~n_{c\bar{c}}(b) .
\label{eq:4.10}
\end{equation}
This cross section can be larger than 
the total $pp$ cross section, because it accounts for multiparton
interactions. In particular, using  $\sigma_{c\bar{c}}$, the contribution of 
charm producing processes to the total cross section defined above, the ratio 
$\sigma_{c\bar{c}\rm incl}/\sigma_{c\bar{c}}$
gives the average charm 
multiplicity.

Notice that here we consider only independent production of $c\bar{c}$ pairs, so 
 that from each pair of colliding partons it results only one $c\bar{c}$ pair, 
and we neglect coherent production of multiple $c\bar{c}$ pairs in
2$\to$4, 2$\to$6, etc.\ processes. This will underestimate the charm production
cross section.

We assume in the following that the partonic distributions $f_i(x,\mu_F^2,b)$
factorize as
\begin{equation}
  f_i(x,\mu_F^2,b) = f_i(x,\mu_F^2) \rho_i(b) ,
\label{eq:4.11}
\end{equation}
where $f_i(x,\mu_F^2)$ is the usual parton distribution function, and
$\rho_i(b)$ is the probability density of finding a parton in the area $d^2b$
at impact parameter $b$. We normalize $\rho_i(b)$ to $\int d^2 b ~\rho_i(b) =
1$, to maintain the usual normalization $\int dx ~x f_i(x) = 1$. The
factorization in Eq.~(\ref{eq:4.11}) is consistent with the usual parton
picture and with our assumption of no parton-parton correlations.

The mean number of $ij\to c\bar{c}$ scatterings at impact parameter $b$ then
becomes
\begin{equation}
  n_{c\bar{c}}(b) = \sum_{ij} a_{ij}(b) \sigma_{QCD}(ij\to c\bar{c}),
\label{eq:4.12}
\end{equation}
where
\begin{equation}
  a_{ij}(b) = \int d^2b' \rho_i(b') \rho_j(b+b')
\label{eq:4.13}
\end{equation}
is an overlap integral, and $ \sigma_{QCD}(ij\to c\bar{c}) $ is the QCD
parton-parton cross section for $ij\to c\bar{c}$, as in Eq.~(\ref{eq:4.4}). 
From
the normalization of $\rho_i(b)$ it follows that $\int d^2b ~a_{ij}(b) =1$ for
every $i,j$. Hence from Eqs.~(\ref{eq:4.10}) and~(\ref{eq:4.12}) we find
\begin{equation}
  \sigma_{c\bar{c}\rm incl} = \sigma_{QCD},
\label{eq:4.14}
\end{equation}
where $\sigma_{QCD}$,  given is 
Eq.~(\ref{eq:4.3}),  is the
charm production cross section calculated within QCD.  This justifies our use
of $\sigma_{QCD}$ as $\sigma_{c\bar{c}\rm incl}$ in the calculation of 
the atmospheric fluxes. 

The way in which we use $\sigma_{c\bar{c}\rm incl}$ in our simulation is as 
follows. We input only one $c\bar{c}$ pair per $pp$ collision at a given energy
 $E$, and multiply  the outcome by
$\sigma_{c\bar{c}\rm incl}$, which includes the $c\bar{c}$ multiplicity. 
We make, therefore, the following  approximation in the kinematics of the 
$c\bar{c}$ pairs produced in the same $pp$ interaction.
Even if in a
real multiparton collision the energy available to the second and other
$c\bar{c}$ pairs is smaller than $E$, we are neglecting this difference. This
is a very good approximation because the fraction of center of mass energy that
goes into a $c\bar{c}$ pair is of the order of $\sqrt{\hat{s}/s} \sim
\sqrt{10{\rm~ GeV}/E} \ll 1$ at the high energies we are concerned with.

Although we have explicitly proven Eq.~(\ref{eq:4.14}) in the absence of
parton--parton correlations, the same result can be obtained when
correlations are present (see sect. V of Ref. \cite{Treleani} and references 
therein).
What is proven even in the presence of correlations is that the pQCD single 
scattering
cross section $\sigma_{QCD}$ is equal to the average number of parton-parton 
collisions,
call it $<N>$, multiplied by the contribution of $c\bar{c}$ producing events to 
the total
cross section (the cross-section $\sigma_{c\bar{c}}$ defined above), namely 
$\sigma_{QCD} = <N> \sigma_{c\bar{c}}$ (while the inclusive cross section is
equal to  the average multiplicity of $c\bar{c}$ pairs
 multiplied by $\sigma_{c\bar{c}}$).   $<N>$ may in general contain 
contributions from two types of collisions. One type consists of collisions of  
pairs of partons (consisting of one parton from each interacting nucleon) which 
interact only once at different points of the transverse plane. Each collision 
of this type results in our case in one $c\bar{c}$ pair produced. The second 
type consists of rescatterings in which one parton of one of the nucleons and 
its interaction products 
interact with several partons of the other nucleon. In  interactions of this 
type, which are much rarer than the first ones, the number of $c\bar{c}$ pairs 
produced not necessarily coincides with the number of collisions. If 
rescatterings 
can be neglected, then $<N>$ is the average number of $c\bar{c}$ pairs produced  
and $\sigma_{QCD}$ is the inclusive $c\bar{c}$ production cross section as 
stated in Eq.~(\ref{eq:4.14}). Otherwise small rescattering corrections, 
to our knowledge not yet calculated \cite{Treleani2}, are necessary (rescatterings 
would also modify the energy spectrum of the particles produced).

\section{Uncertainties due to cascade simulation, 
parameters of charm production model and choice of PDF's} 
 
\label{sect:charm4} 
 
In our first paper (GGV1) we considered the uncertainties related to the
cascade generation in PYTHIA. 
There we tried different modes of cascade generation,
different options allowed by PYTHIA in the various stages of parton showering,
hadronization, interactions and decays, etc., without noticing substantial
changes in the final results (differing at most by $10~\%$). These
uncertainties are however very difficult to quantify, due to the nature of the
PYTHIA routines. Since these uncertainties are small, we neglect them in this
analysis and continue to use PYTHIA with the options described in GGV1 as our
main choice for the simulation: `single' mode with showering, `independent'
fragmentation, interactions and semileptonic decays according to TIG.

In our `single' mode we enter only one $c$ quark in the particle list of PYTHIA, 
and we multiply the result by a factor of $2$ to account for the initial 
$\bar{c}$ quark. 
PYTHIA performs the showering, standard independent fragmentation, and follows 
all 
the interactions and decays using default parameters and options. In GGV1 we 
have 
argued that this `single' approach is equivalent to what we called `double' 
mode, 
in which both $c\bar{c}$ partons are placed in the initial event list, in the 
first 
step of a standard cascade evolution.
The `single' option is chosen thus, because it reduces considerably the 
computing time.
 
Important sources of uncertainty are contained in our charm production
model, which is NLO pQCD as implemented in the MNR program \cite{MNR},
calibrated at low energies.
 
The calibration procedure consisted in the following: 
 
$\bullet $ choosing a PDF set from those available and fixing the related value
of $ \Lambda _{QCD}$;\footnote{We note that $\Lambda _{QCD}$ can be chosen in
  the MNR program independently of the PDF and therefore can constitute an
  additional independent parameter of our model. We have opted however to
  choose the value of $\Lambda _{QCD}$ assumed in the PDF set being used, for
  consistency.}
 
$\bullet$ choosing $m_{c}$, $\mu_{F}$ and $\mu_{R}$, which are the charm mass,
the factorization scale and the renormalization scale respectively, so as to
fit simultaneously both the total and differential cross sections from existing
fixed target charm production experiments \cite{alves1,alves2} at the energy
of 250 GeV, without additional normalization factors;
 
$\bullet$ checking that the total cross section generated after the previous
choices fits reasonably well the other existing experimental points for fixed
target charm production experiments \cite{fmnr}.
 
Besides the choice of the  PDF set,
 our procedure has the freedom to choose
reasonable values of the three parameters $m_{c}$, $\mu_{F}$,
 and $\mu_{R}$ so as to
fit the experimental data. In GGV1 and GGV2 we made the standard
choice  \cite{MNR,fmnr}  of
\begin{equation} 
\mu_{F}= 2m_{T}, \quad \mu_{R} = m_{T},  
\label{eq:4.15} 
\end{equation} 
where $m_{T} = \sqrt{p_{T}^{2} + m_{c}^{2}}$ is the transverse mass. The
values of the charm mass are taken slightly different for each PDF set,
namely:
\begin{eqnarray} 
m_{c}&=&1.185{\rm ~GeV} \quad\hbox{for MRS R1,}  \label{eq:4.16} \\ 
m_{c}&=&1.310{\rm ~GeV} \quad\hbox{for MRS R2,}  \label{eq:4.17} \\ 
m_{c}&=&1.270{\rm ~GeV} \quad\hbox{for CTEQ 4M,}  \label{eq:4.18} \\ 
m_{c}&=&1.250{\rm ~GeV} \quad\hbox{for MRST.}  \label{eq:4.19} 
\end{eqnarray} 

Here we explore the changes induced  in cross sections and fluxes at high
energies by different choices of $m_{c}$, $\mu_{F}$,
 and $\mu_{R}$ which fulfil our calibration requirements.

We have performed this  analysis with the most recent PDF set: the
MRST \cite{MRST} (other PDF's give similar results).
At first we fix  $\lambda = 0$ and then we examine other values of $\lambda$.
We note that the calibration procedure described above is
independent of $\lambda$ because it involves only relatively low energies,
where the low $x$ extrapolation is not an issue.

\subsection{MRST $\mathbf \protect\lambda = 0$: fluxes}
\label{subsect:0fluxes4} 
 
We considered the $\lambda = 0$ case because it is the most significant for the
evaluation of the uncertainties in the spectral indices, as it will be clear in
the next subsection. We have considered the following reasonable ranges of the
parameters
\begin{eqnarray} 
1.1{\rm ~GeV} \leq m_{c} \leq 1.4{\rm ~GeV},  \label{eq:4.20} \\ 
0.5~m_{T} \leq \mu_{F} \leq 2~m_{T},  \label{eq:4.21} \\ 
0.5~m_{T} \leq \mu_{R} \leq 2~m_{T},  \label{eq:4.22} 
\end{eqnarray} 
where the bounds on $m_c$ come from the 1998 Review of Particle Physics 
\cite{RPP98}, while those for $\mu_{F}$ and $\mu_{R}$ are those used in the 
existing
literature \cite{MNR,fmnr}.
 
Within these ranges we have looked for values of the three parameters capable
of reproducing the experimental data in our calibration procedure described
before. Table~\ref{table:1} summarizes the different sets of parameters: we
have varied the charm mass through the values $m_c = 1.1$, 1.2, 1.25, 1.3, 1.4
GeV ($m_c=1.25$ GeV was our previous optimal choice for MRST in
Eq.~(\ref{eq:4.19})) and then, for each value of $m_c$, we have found values of 
the factorization and
renormalization scales that reproduce the experimental total cross
section $\sigma_{c\bar c} = 13.5 \pm2.2$ $\mu$b at 250 GeV \cite{alves1}. In
particular, for each value of $m_c$, we took
$\mu_{F} = m_{T}/2$, $m_{T}$, $2m_{T}$ and then, for each $m_c$, $\mu_{F}$ pair, 
found the value of
$\mu_{R}$ which best fits the data (see Table~\ref{table:1}).
 
We have also checked that these choices give good fits to the differential,
besides the total, cross sections at 250 GeV \cite{alves2}, without additional
normalization factors, as done for the original choice of parameters in GGV1.
For $m_c = 1.1$ GeV we had to choose values of $\mu_{R}$ slightly outside the
range in Eq.~(\ref{eq:4.22}) (but we have kept these values in our analysis
anyway).
 
For all the sets of parameters in Table~\ref{table:1} we have run our full
simulations for the MRST, $\lambda = 0$ case and the results are described in
Figs.\ 2-4.
 
In Figs.\ 2a and 2b we show the resulting total charm production cross section
$\sigma_{c\bar c}$ for all of the fifteen sets of parameters in 
Table~\ref{table:1}, together with recent
experimental data (from Table 1 of Ref.~\cite{fmnr}, where all the data for $
pp$ or $pN$ collisions have been transformed into a $\sigma_{c\bar c}$ cross
section following the procedure described in GGV1). Fig.\ 2b is  an
enlargement of the region of Fig.\ 2a containing the experimental data.
 
In Fig.\ 2a we see the spread of the cross sections, which is more
than one order of magnitude at  $ 10^{11}$ GeV. 
Above $250~GeV$, one can clearly distinguish  three 
``bands'' of increasing cross sections for  
$\mu_{F} = m_{T}/2$, $m_{T}$ and $2m_{T}$. Within each ``band''
the cross sections
increase with increasing values of $m_c$ (and correspondingly smaller values of 
$\mu_{R}$), in Table~\ref{table:1}. 
Our standard choice ($m_c = 1.25$ GeV, $\mu_{F} = 2m_{T}$, $\mu_{R}
= m_{T}$) proves to be one of the highest cross sections we obtain.
 
In Fig.\ 2b we see better how all of these cross sections verify our calibration
procedure. They  pass through the point at 250 GeV \cite{alves1}, agree with the 
point at 400 GeV \cite{aguilar88} and
disagree with the very low experimental point at 200 GeV \cite{barlag88}. 
The lower values of $\mu_{F} = m_T/2$ and
$m_{T}$ fit better the lowest experimental point at 800 GeV \cite{ammar88}, 
while the higher value of $\mu_{F} = 2m_{T}$ fits better the upper
point at 800 GeV \cite{kodama91}.
 
We believe that the spread of the total cross sections shown in Fig.\ 2a
 provides a reasonable estimate of the uncertainty of our
charm production model at  fixed $\lambda$. 
Since our standard choice of parameters ($m_c = 1.25$ GeV, $\mu_{F} = 2m_{T}$, 
$\mu_{R} = m_{T}$) gives one of the highest cross sections
 (in better agreement  with the more recent value of the cross section at
800 GeV \cite{kodama91}), the uncertainty band should be added under each of the 
cross section curves calculated with our standard choice of parameters (like the 
curves shown in Fig.~1 of GGV2).
 
Fig.\ 3 illustrates the corresponding spread of the final prompt fluxes.
 Although our results are for the
 MRST PDF's extrapolated with $\lambda = 0$ (the value of 
 $\lambda$ which gives the lowest  fluxes) similar spreads result from other
 PDF's and  $\lambda$'s. We show the flux of muons;
the fluxes of muon-neutrinos and electron-neutrinos are essentially the
same.
 
Similarly to what happens with cross sections in Fig.\ 2, the fluxes in Fig.\ 3
increase with $\mu_{F} = m_{T}/2$, $m_{T}$ and $2m_{T}$ and, within each band, 
they increase
with increasing $m_c$ (and correspondingly smaller values of $\mu_{R}$), in 
Table~\ref{table:1}. 
At energies around $10^6$ GeV the total uncertainty is almost one
order of magnitude and decreases slightly for higher energies. If we would
 decide to work only with $\mu_F = 2m_{T}$  (which fits the 
experimental measurement at 800 GeV with the highest cross section), 
the uncertainty would be
greatly reduced: the fluxes in this rather narrow band  differ by
less than 40\%. We observe that the flux calculated with
our standard choice of parameters ($ m_c = 1.25$ GeV, $\mu_{F} = 2m_{T}$,
$\mu_{R} = m_{T}$) is almost the highest, as was the case for the corresponding
cross section in Fig.~2.
 
In Fig.\ 3 we also indicate the conventional and prompt fluxes from TIG; we
notice that the TIG prompt flux is within our band of uncertainty,
which is reasonable since TIG used a low $\lambda=0.08$ value for their
predictions (see the discussions in GGV1 and GGV2).
 
\subsection{MRST $\mathbf \protect\lambda = 0$: spectral index} 
\label{subsect:0index4}

In our previous paper GGV2, we pointed out that 
an experimental measurement of the
slope of the atmospheric lepton fluxes at energies where the prompt component 
dominates over
the conventional one, might give information on the value of $\lambda $,
the slope of the gluon PDF at small $x$.  The 
best flux for this measurement is that of down-going muons, because
the prompt neutrinos have first to convert into muons or electrons through
a charged current interaction in order to be detectable in a neutrino
telescope. 

In this section and in the following two we consider the
uncertainties in our method to determine $\lambda $. In this section we examine 
those
coming from the charm production model, in Sect.~\ref{sect:lambda4} those 
related to the non linearity of our model, and in Sect.~\ref{sect:composition4} 
those
coming from the unknown composition of the cosmic rays at high energies.
 
The slope of the fluxes or spectral index is $\alpha_{\ell}(E_{\ell})
= - \partial \ln \phi_{\ell}(E_{\ell}) / \partial \ln E_{\ell} $, with $\ell =
\mu^\pm, \nu_\mu + \bar {\nu_\mu}$ or $\nu_e+ \bar {\nu_e}$. In other words,
the final lepton fluxes are 
\begin{equation} 
\phi_{\ell}(E_{\ell}) \propto E_{\ell}^{-\alpha_{\ell}(E_{\ell})}. 
 \label{eq:4.23} 
\end{equation} 
In GGV2 we found a simple linear dependence of
$\alpha_{\ell}$ on $\lambda$, namely
\begin{equation} 
\alpha_{\ell}(E_{\ell}) = b_{\ell}(E_{\ell},\gamma,\lambda) - \lambda \simeq 
b_{\ell}(E_{\ell}) - \lambda,  
\label{eq:4.24} 
\end{equation} 
where $b_{\ell}(E_{\ell})$ is an energy dependent coefficient 
evaluated using our simulation with $\lambda = 0$ and fixed $\gamma$.
 As argued in GGV2 (cfr.\ Eqs.~(35) and~(36) therein), the coefficient 
$b_{\ell}(E_{\ell},\gamma,\lambda)$ depends mildly on $\lambda$ and can be well 
approximated by its value for $\lambda=0$ (see Sect.~\ref{sect:lambda4}).
The coefficient $b_{\ell}(E_{\ell},\gamma,\lambda)$ depends almost linearly on 
$\gamma$, the spectral index of the 
primary cosmic rays. We recall that the equivalent nucleon flux for primary 
cosmic rays is expressed as
\begin{equation}
\phi_{N}(E) \propto E^{-\gamma-1}.
\label{eq:4.24.1}
\end{equation}
The linear dependence of $b_{\ell}(E_{\ell},\gamma,\lambda)$ on $\gamma$ can be 
written as
\begin{equation} 
b_{\ell} (E_{\ell},\gamma,\lambda) = {\bar b}_{\ell}(E_{\ell}, \gamma, \lambda) 
+ \gamma,  
\label{eq:4.25} 
\end{equation} 
where ${\bar b}_{\ell}(E_{\ell}, \gamma, \lambda)$ depends
 mildly on $\lambda$ and $\gamma$ \footnote{We have included in ${\bar 
b}_{\ell}$ the $+1$ term coming from the $-1$ in the exponent of 
Eq.~(\ref{eq:4.24.1}).}, as we will prove in Sect.~\ref{sect:lambda4} and 
Sect.~\ref{sect:composition4}, respectively.

Given $b_{\ell }(E_\ell)$ from our model, an
experimental measurement of $\alpha _{\ell }$ at energy $E_{\ell }$ would
immediately give $ \lambda $ corresponding to a value of $x\simeq {\rm
  GeV}/E_{\ell }$, as we discussed in GGV2. A measurement 
  at $E_{\ell}\simeq 10^{6}$ GeV = 1 PeV would give $\lambda $ at $x\simeq 
10^{-6}$,
a value of $x$ unattainable by present experiments.

For the time being we keep fixed the value of $\gamma$ ($\gamma=1.7$ below the 
knee, and $\gamma=2.0$ above the knee, as in GGV1 and GGV2); only in 
Sect.~\ref{sect:composition4} we will consider changes in the value of $\gamma$.
 
The feasibility of a measurement of $\lambda$ depends, therefore, on the
uncertainties in $b_{\ell}(E_{\ell})$. Here we discuss those coming from
 the charm production model.
 
Fig.~4 shows the $-b_{\mu }$ corresponding to the fluxes of Fig.~3 
as functions of $E_\mu$. In the region of interest 
$E_{\mu}\gtrsim 10^{5}$--$10^{6}$ GeV,  the values of $ -b_{\mu}$
within each ``band'' decrease with increasing
$m_{c}$ (and correspondingly smaller values of $\mu_{R}$), in 
Table~\ref{table:1}.
 
The  spread of $b_{\mu}$ due to the choice of $\mu_F$ , $\mu_R$ and $m_c$
is  $\Delta b_{\mu}
\simeq 0.1$, or $\Delta b_{\mu}/ b_{\mu} \simeq 0.03$,  much smaller 
than the uncertainty $\Delta \phi_{\mu}/ \phi_{\mu} \simeq 10$ of the absolute
flux in Fig.~3. If we would for some reason restrict ourselves to
the $\mu_F = 2~m_{T}$ band, the
uncertainty on $b_{\mu}$ would become even smaller, $\Delta b_{\mu} \simeq 0.03 
$. We will refer to this error as $\Delta b_{par}$ in the following, as it is 
related to the choice of parameters in the charm production model, and consider 
half of the spread in Fig.~4 to evaluate it. Therefore

\begin{equation} 
\Delta b_{par} \simeq 0.05~(0.015),  
\label{eq:4.26} 
\end{equation} 
where the value in parenthesis corresponds to the $\mu_F=2m_T$ band.

\subsection{MRST $\mathbf \protect\lambda = \protect\lambda$(T)}
\label{subsect:Tfluxes4} 
 
So far we used $\lambda = 0$ only. This case determines the uncertainty of the
$b_{\ell}(E_{\ell})$ function which enters in the determination of
$\lambda$ through the atmospheric muon fluxes.
 
Here we study  an ``intermediate'' value of $\lambda$.
We continue to use the MRST PDF,
but with the value of $\lambda = \lambda(T)$
 given by the slope of the lowest tabulated 
 value of $x$ (see GGV2 for more explanations). This value depends on $Q^2$ and
 is about 0.2-0.3.
 
We repeat the same analysis of subsection \ref{subsect:0fluxes4}. However,
for simplicity, we report
the results  for four selected sets of values for the parameters in Table~\ref
{table:1}. The first set ($m_c = 1.1$ GeV, $\mu_F = 0.5~m_{T}$, $\mu_R
= 2.53~m_{T}$) gives a lower bound for the fluxes.
The second set ($m_c = 1.4$ 
GeV,
$\mu_F = 2m_{T}$, $\mu_R = 0.61~m_{T}$) 
gives an upper bound for the
fluxes. The remaining two sets are cases in the $\mu_F = 2m_{T}$ ``band''.
 
The results are plotted in Fig.~5. 
The general features of Fig.~5 coincide with those of Fig.~3,
 except for an overall increase in
all the fluxes due to the larger value of $\lambda$. The total spread of the
fluxes given by the two limiting cases, as well as the spread within
the narrower $\mu_F = 2m_T$ band, are  comparable to those found for
$\lambda = 0$. As in Fig.~3, our standard choice of parameters
($m_c = 1.25~{\rm GeV},~\mu_{F} = 2m_{T},~\mu_{R} = 1.0~m_{T}$) yields
almost the highest flux.
 
We  conclude that similar features would be obtained  for other values
of $\lambda$: our ``standard choice'' flux would be almost the highest in a
band of uncertainty whose width is similar for all values of $\lambda$.
The fluxes in the uncertainty band of Fig.~5 are
consistent with older predictions (see GGV1 and references therein) and with
the prediction by L.~Pasquali et al.~\cite{PRS}.
 
\subsection{Other PDF's} 
\label{subsect:PDF4}
 
Another source of uncertainty for the final fluxes and spectral indices is the
choice of the PDF set. As  in GGV2, we consider here four recent sets of PDF's:
MRS R1-R2 \cite{MRS1}, CTEQ 4M \cite{CTEQ} and MRST \cite{MRST},
with the standard choice of parameters of 
Eqs.~(\ref{eq:4.15}),(\ref{eq:4.16}),(\ref{eq:4.17}),
(\ref{eq:4.18}),(\ref{eq:4.19}). 
 
Figs.~6a and~6b show the muon fluxes (top panels) and
spectral indices (bottom panels) for the two limiting cases of $\lambda =0$
(Fig.~6a) and $\lambda=0.5$ (Fig.~6b). In both cases the $\mu$ fluxes show 
at most a $30-50\%$ variation depending on the PDF used. The uncertainty 
in the spectral indices  for
$E_{\mu}\gtrsim 10^{5}-10^{6}$ GeV is $\Delta b_{\mu}\lesssim 0.02$, or $\Delta
b_{\mu}/ b_{\mu}\lesssim 0.01$. This error will be denoted as $\Delta b_{PDF}$ 
in the following, namely 
(again dividing the spread by $2$)

\begin{equation} 
\Delta b_{PDF} \simeq 0.01.  
\label{eq:4.27} 
\end{equation} 

These uncertainties, related to the PDF's, are smaller that those due to
the choices of mass scales (see Figs.~3-4).
We  conclude that, provided different
PDF's are calibrated in a similar way (i.e. same values of $\mu _{F}$, $\mu
_{R}$ and $m_{c}$, chosen to fit the experimental data of our calibration),
the final fluxes and spectral indices are very similar. 
The main source of uncertainty resides therefore in the choice of the mass
parameters, rather than the adoption of a certain PDF set.
 
\section{Determination of $\protect\lambda$ with neutrino telescopes} 
\label{sect:lambda4} 
 
In GGV2 we have given a detailed analysis of the dependence of the final fluxes
and spectral indices on $\lambda $ for different PDF's. 
In this section we show that the spread in our results due to $\lambda$ 
is larger than the one due to the choice of 
$\mu _{F}$, $\mu_{R}$, $m_{c}$ and of the PDF set, analyzed in the previous 
section.
 This is good news for the possibility of measuring $\lambda$, since the 
 spread in $\alpha_\mu$, due to different $\lambda$'s, is the signal we want to 
detect, while the spread 
 due to other factors constitutes the theoretical error of this measurement.

Figs.~7-10 show how the $\mu$ flux and its spectral index depend on $\lambda$.
 We used MRST with variable $\lambda =0,~0.1,~0.2,~0.3,~0.4,~0.5$
 and our standard choice of parameters ($m_c = 1.25~{\rm
  GeV}$, $\mu_F = 2m_{T}$, $\mu_R = 1.0~m_{T}$).

Fig.~7 contains the differential muon fluxes. At the highest energies 
the $\mu$ fluxes are spread over almost two orders of magnitude.
To each of the curves in this plot we need to assign a band of uncertainty
of about one order of magnitude coming from the choice of the PDF and of the
parameters $m_c$, $\mu_F$, and $\mu_R$ (see Fig. 3). Thus the curves become 
entirely superposed with each other. This makes it
impossible to derive the value of $\lambda$ from an experimental measurement of 
the
absolute level of the fluxes. However, the uncertainties in the spectral 
index of these prompt muons are much smaller and a determination 
of $\lambda$ becomes possible using the slope of the muon fluxes instead of
their absolute level.
 
Fig.~8 shows the $E^2$-weighted integrated fluxes as functions of the muon
energy. The slant lines indicate the number
of particles  traversing a km$^3$ detector over a $2\pi$ sr solid angle.
Even for the highest predicted fluxes, less than 1 particle per year will 
traverse the km$^3$ detector for energies
above $10^8$ GeV. Moreover, while prompt muons can be detected directly,
prompt neutrinos have first to convert into charged
leptons before being detected. The smallness of the charged current
interaction effecting the conversion considerably lowers the detection rate of 
neutrinos.
 Therefore, the slope of the charm component 
 of the atmospheric leptons can be studied in
 neutrino telescopes only
using  atmospheric muons coming from above the horizon, and only
 in a narrow range of energies, between a lower limit of $E_{\mu}\simeq
10^{5}-10^{6}$ GeV, above which the prompt component dominates over the
 conventional one, and an upper limit of $E_{\mu}\simeq 10^{7}-10^{8}$ GeV, 
above which the detection rates become negligible.

In practice, the spectral index of the prompt muon flux may be estimated by
the difference of two integrated muon fluxes above two different energies, e.g.
$10^6$ and $10^7$ GeV.
 
Figs.~9, 10 prove the validity in our model of Eq.~(\ref{eq:4.24}), which is
$\alpha_{\ell}(E_{\ell}) = b_{\ell}(E_{\ell}) - \lambda$. In Fig.~9 we plot the
spectral indices $-\alpha _{\ell }(E_{\ell })$ for the different values of
$\lambda $, both as directly calculated with our simulation (solid lines) and
as $-b_{\ell }(E_{\ell })+\lambda $ (dotted lines), where $b_{\ell }(E_{\ell
  })$ is $\alpha_{\ell}$ with $\lambda =0$.  The two almost coincide, in the
interval of interest, $E_{\ell }\gtrsim 10^{6}$ GeV.  Their difference, $\alpha
_{\ell }(E_{\ell })-b_{\ell }(E_{\ell })+\lambda $, given in Fig.~10, is small,
about $\simeq 0.03$ at $E \gtrsim 10^6$ GeV.  This difference stems from the 
mild
dependence of $b_{\ell }(E_{\ell })$ on $\lambda$ and need to be added to the
the other uncertainties evaluated in Sect.~\ref{sect:charm4}. We will refer to 
this error, due to the non linearity in $\lambda$ of Eq.~(\ref{eq:4.24}), as

\begin{equation} 
\Delta b_{non-lin} \simeq 0.015,  
\label{eq:4.28} 
\end{equation} 
which again is evaluated dividing by $2$ the spread in Fig.~10.

We see in Fig.~9 that
$\Delta \lambda \sim \Delta \alpha$, therefore we would need an uncertainty in 
the
spectral index of order $0.1$ to determine $\lambda$ with the same accuracy. We 
will show now that this is roughly the uncertainty related to our
theoretical model.
 
In fact, we can finally estimate the total uncertainty in the determination
of $\lambda$ coming from our theoretical model (that is, excluding the
uncertainty due to the unknown composition of cosmic rays). We sum together the 
three
spreads of $b_{\ell}(E_{\ell})$ previously calculated in Eqs.~(\ref{eq:4.26}), 
(\ref{eq:4.27}) and (\ref{eq:4.28}), to obtain the final
uncertainty \footnote{We summed the errors linearly. Summing in quadrature
  would give $(\Delta \lambda)_{charm} \simeq (\Delta b_\mu)_{charm} \simeq 
0.053~(0.023)$.} from the charm production model,

\begin{equation} 
(\Delta \lambda)_{charm} \simeq (\Delta b_\mu)_{charm} \simeq 0.075~(0.04),  
\label{eq:4.29} 
\end{equation} 
where the number in parenthesis corresponds to fixing $ \mu_F =
2m_{T}$  in the charm model.
 
If the theoretical uncertainties so far presented would be 
the only ones affecting the determination of $\lambda$ 
through a measurement of the slope of the down-going muon 
flux, we could expect to get to know $\lambda$ with an uncertainty 
of about $\Delta \lambda \sim 0.1$. However, even excluding 
experimental uncertainties in the neutrino telescopes themselves,
the uncertainty increases when our ignorance of
the composition of the cosmic rays at high energy is taken into account,
as we show in the following section.

\section{Uncertainty from cosmic ray composition} 
\label{sect:composition4}

The final uncertainty we consider in the determination of $\lambda$ comes from
the poorly known elemental composition of the high energy cosmic rays. 

The spectral index of the cosmic rays $\gamma$ enters almost linearly in the
slope of the atmospheric leptons. From Eqs.~(\ref{eq:4.24}) and (\ref{eq:4.25}) 
we have 
\begin{equation}
\alpha_{\ell}(E_{\ell}) = 
 {\bar b}_{\ell}(E_{\ell}, \gamma, \lambda) + \gamma - \lambda.
\label{eq:4.30}
\end{equation}
So far we have kept $\gamma$ fixed, thus the uncertainty $\Delta b_{\mu}$ 
calculated in Eq.~(\ref{eq:4.29}) is actually an uncertainty in ${\bar 
b}_{\mu}$.
We are going now to evaluate the uncertainty due to $\gamma$. 

The non--linearity of Eq.~(\ref{eq:4.30}) with respect to $\gamma$ is
mild, as we have argued analytically in Sect.~V of GGV2 and we show here using 
our numerical simulation.  
We have conducted a few trial runs of our simulation simply changing the values 
of $\gamma$ used for the primary flux. We recall from subsection 
\ref{subsect:0index4} that in our model we used $\gamma=1.7,~2.0$, respectively 
below and above the knee at $E=5~10^6~GeV$. We have run our simulation changing 
these values of $\gamma$ by $\pm 0.1,~\pm 0.2$ \footnote{Notice that these 
values of $\gamma$ are some of the most probable values (see Fig.~13).}, both 
above and below the knee, to see the error produced when taking ${\bar 
b}_{\ell}$ computed at fixed $\gamma$ (our usual values) in Eq.~(\ref{eq:4.30}) 
and thus leaving a pure linear dependence on $\gamma$. We used the MRST PDF, 
with $\lambda=0$, but similar results are obtained with other PDF's and 
$\lambda$'s.

In Fig.~11a we plot the spectral index $-\alpha_{\ell}(E_{\ell})$ for the 
different values of $\gamma$, both as directly calculated with our simulation 
(solid lines) and as $-{\bar b}_{\ell}(E_{\ell}; \gamma=1.7,~2.0; 
\lambda=0)-\gamma$ (dotted lines), i.e. using our standard values for $\gamma$ 
of the primary flux and adding an increment in $\gamma$ equal to $\pm 0.1,~\pm 
0.2$. In this way the ``central value'' of these spectral indices corresponds to 
the $\lambda=0$ case of Fig.~9. We can see that the dotted and the solid lines 
almost coincide, especially in the interval of interest for $E_{\ell }\gtrsim 
10^{5}-10^{6}$ GeV, proving the validity of Eq.~(\ref{eq:4.30}). The uncertainty 
in ${\bar b}_{\ell}$ due to this non-linearity, that we call $\Delta 
\gamma_{non-lin}$, evaluated in terms of the difference $\alpha_{\ell}-{\bar 
b}_{\ell}-\gamma$, is plotted in Fig.~11b and, in the energy range of interest, 
is 

\begin{equation} 
\Delta \gamma_{non-lin} \simeq 0.02.  
\label{eq:4.31} 
\end{equation} 

We will now consider the error due to the poorly known elemental composition of 
the high energy cosmic rays. 
Concerning charm production, the relevant cosmic ray flux to be considered is 
the
equivalent flux of nucleons impinging on the atmosphere. For a given cosmic ray
flux, the equivalent flux of nucleons $\phi_{eq}(E_N)$ depends in general on
the composition of the cosmic rays. Nuclei with atomic number $A$ and energy
$E_A$, coming with a flux $\phi_A(E_A)$, contribute an amount $A \phi_A(AE_N)$
to the equivalent flux of nucleons at energy $E_N = E_A/A$. So in total
\begin{equation}
\phi_{eq}(E_N) = \sum_A A \phi_A(A E_N) .
\label{eq:4.32}
\end{equation}
The uncertainties in the equivalent nucleon flux arise from the poorly known
composition of cosmic rays in the energy range above the so-called knee, $E_A
\sim 10^6$ GeV.

The actual $\gamma$ that enters into our proposed determination of $\lambda$ is
the spectral index of the equivalent nucleon flux $\gamma_{eq}$, the equivalent
cosmic rays spectral index for short. The equivalent nucleon flux is written as 
$\phi_{eq} \propto E_N^{-\gamma_{eq}-1}$, so that the spectral index 
$\gamma_{eq}$ is given by
\begin{equation}
  \gamma_{eq} +1= - \frac{E_N}{\phi_{eq}} \, \frac{\partial \phi_{eq}}{\partial 
E_N}
  = \frac{1}{\phi_{eq}} \sum_A A \phi_A (\gamma_A+1) ,
\label{eq:4.33}
\end{equation}
where $\gamma_A$ is the spectral index of the component 
of atomic number $A$, i.e. $\phi_A(E_A) = k_A E_A^{-\gamma_A-1}$.

We have calculated $\phi_{eq}$ and $\gamma_{eq}$ using the experimental data of 
JACEE~\cite{JACEE}, CASA-MIA~\cite{CASA-MIA}, HEGRA~\cite{HEGRA}, and the data 
collected by Biermann et al, in Table 1 of Ref.~\cite{BIERMANN}, each with
their respective compositions.
Figs.~12 and~13 show the $\phi_{eq}$ and the $\gamma_{eq}$ so obtained. 
Only the data of CASA-MIA~\cite{CASA-MIA} and HEGRA~\cite{HEGRA} reach 
energies $E_N \lesssim 10^8$ GeV, so we have not extended our analysis beyond
$10^8$ GeV.

We have calculated the error band associated to $\gamma_{eq}$ in
two different ways,
because of the different parametrization of the composition
 used in Refs. \cite{JACEE} to \cite{BIERMANN}.
Refs. \cite{JACEE,BIERMANN} give  separate power law fits to the spectrum of 
each
cosmic ray component,
\begin{equation}
  \phi_A(E_A) = k_A E_A^{-\gamma_A-1} ,
\label{eq:4.34}
\end{equation}
where the parameters $k_A$ and $\gamma_A$ have errors $\Delta k_A$ and $\Delta
\gamma_A$. Standard propagation of errors gives, in this case,
\begin{equation}
  \Delta \phi_{eq} = \left\{  \sum_A A^2 \phi_A^2 
    \left[ \Biggl(\frac{ \Delta k_A}{k_A}\Biggr)^2 + 
  \Bigl( \ln(A E_N) \Delta \gamma_A \Bigr)^2
  \right] \right\}^{1/2}
\label{eq:4.35}
\end{equation}
and
\begin{equation}
  \Delta \gamma_{eq} = \left\{
    \sum_A \frac{A^2 \phi_A^2}{\phi_{eq}^2} 
    \left[ (\gamma_A-\gamma_{eq})^2 \Biggl(\frac{\Delta k_A}{k_A}\Biggr)^2  + 
     \Bigl[ 1-(\gamma_A-\gamma_{eq})\ln(A E_N)\Bigr]^2
       \Bigl( \Delta \gamma_A \Bigr)^2 
    \right] \right\}^{1/2} ,
\label{eq:4.36}
\end{equation}
where $\phi_A$ is evaluated at $E_A=A E_N$.

Refs.~\cite{CASA-MIA,HEGRA}, give a power law fit to the total particle flux 
\begin{equation}
  \phi(E_A) = k E_A^{-\gamma-1}
\label{eq:4.37} 
\end{equation}
and a composition ratio $r_A(E_A)$ 
in terms of which
\begin{equation}
  \phi_A(E_A) = r_A(E_A) \, \phi(E_A) .
\label{eq:4.38}
\end{equation}
These experiments distinguish only between a light and a heavy component.
 We assign atomic number 1 to the light component and 56 to the heavy one
(which we call ``iron'').  Here $k$, $\gamma$, and $r_A$ have errors $\Delta k$,
$\Delta \gamma$, and $\Delta r_A$, 
respectively. The equivalent nucleon flux is still given 
by Eq.~(\ref{eq:4.32}), while standard propagation of errors
gives in this case
\begin{equation}
  \Delta \phi_{eq} = \left\{  \sum_A A^2 \phi_A^2
   \Biggl(\frac{ \Delta r_A}{r_A}\Biggr)^2 +
     \phi_{eq}^2 \Biggl(\frac{ \Delta k}{k}\Biggr)^2 + 
  \Bigl[ \sum_A A \phi_A \ln(A E_N) \Delta \gamma_A \Bigr]^2
  \right\}^{1/2},
\label{eq:4.39}
\end{equation}
We omit the much longer expression for $\Delta \gamma_{eq}$.
For simplicity, we have  neglected the error coming from
the energy dependence of $r_A$, which we expect to be much smaller than the 
others.
In Fig.~12 we show the equivalent nucleon flux $\phi_{eq}$. It is clear from
the figure that the systematic uncertainties dominate,
 with spreads between different experiments of up to
a factor of 4. 

The uncertainties in the equivalent spectral index $\gamma_{eq}$ are smaller,
as can be seen in Fig.~13, where only HEGRA and CASA-MIA extend to the energy 
region above the knee which is important to us.

We can consider, for example, an energy $E_N \simeq 10^7~GeV$, which is likely 
to determine the leptonic fluxes at around $E_{\ell} \simeq 10^6~GeV$, energy at 
which we would like to measure $\lambda$ through the spectral index (we recall 
from GGV2 that $E_{\ell} \lesssim~0.1~E_N$).

At this energy $E_N$, from Fig.~13, we may take half the difference between the 
central values of the CASA-MIA and HEGRA data as an indication of the systematic 
uncertainty on $\gamma_{eq}$,

\begin{equation} 
\Delta \gamma_{syst} \simeq 0.1.  
\label{eq:4.40} 
\end{equation} 

Using the CASA-MIA data and the related error band, instead of the very spread 
HEGRA data, we can expect a reasonable statistical uncertainty

\begin{equation} 
\Delta \gamma_{stat} \simeq 0.05.  
\label{eq:4.41} 
\end{equation}
Since $\alpha_{\ell}$ depends
linearly on $\gamma_{eq}$ and $\lambda$, the same uncertainties apply to a 
determination of
$\lambda$. 
The total uncertainty in the determination
of $\lambda$ coming from the unknown composition of cosmic rays is now simply 
the sum of Eqs.~(\ref{eq:4.31}), (\ref{eq:4.40}) and (\ref{eq:4.41}),
\begin{equation} 
(\Delta \lambda)_{comp} \simeq (\Delta \gamma_{eq})_{comp} \simeq 0.17,  
\label{eq:4.42} 
\end{equation} 
if summing the errors linearly, or
\begin{equation} 
(\Delta \lambda)_{comp} \simeq (\Delta \gamma_{eq})_{comp} \simeq 0.11,  
\label{eq:4.43} 
\end{equation} 
if we sum them in quadrature.

Finally, we can now combine all the uncertainties together, to compute the 
overall theoretical error in the determination of $\lambda$ with neutrino 
telescopes.
From Eqs.~(\ref{eq:4.26}), (\ref{eq:4.27}), (\ref{eq:4.28}), (\ref{eq:4.31}), 
(\ref{eq:4.40}), and (\ref{eq:4.41}) we obtain
\begin{equation}
\Delta \lambda \simeq  0.25~~(0.21) 
\label{eq:4.44}
\end{equation}
if summing errors linearly,
or
\begin{equation}
\Delta \lambda \simeq  0.13~~(0.11), 
\label{eq:4.45}
\end{equation}
if summing in quadrature, where the numbers in parenthesis correspond to the $ 
\mu_F = 2m_{T}$ ``band'' in the charm model.

\section{Conclusions} 
\label{sect:conclusions4} 

We have examined in detail the possibility of determining the slope $\lambda$ of 
the gluon PDF, at momentum fraction $x \lesssim 10^{-5}$, not reachable in 
laboratories, through the measurement in neutrino telescopes of the slope of 
down-going muon fluxes at $E_{\mu} \simeq x^{-1}~GeV$.

The method we are proposing may reasonably well reach 
$x \simeq 10^{-7}$, what would require $10~PeV$ in muon energy.
 At this energy, there would still be $50$
events from charm if $\lambda=0.5$ and $10$ events if $\lambda=0$. But the best
measurement could be done between $100~TeV$ and $1~PeV$ of muon energy, i.e. 
between 
$x \simeq 10^{-4}$ and $x \simeq 10^{-6}$. Present data do not go below 
$x \simeq 10^{-5}$ and the Large Hadron Collider (LHC) will not do any better. 
 The reason is that the dominant values of $x$  
in the production of a heavy particle of mass $M$ and 
rapidity $y$ are of order $x\simeq [M \exp{(\pm y)}/ \sqrt{s}]$
(see for example \cite{QCD})
where $\sqrt{s}$ is the center-of-mass energy of the hadron collision. Thus 
the smaller values of $x$ are obtained with the smaller $M$ and larger $y$ for 
fixed
$\sqrt{s}$ ($14~TeV$ at the LHC). Even if exhaustive studies 
of the possible minimum $x$ to be reached at the LHC have not yet been 
carried out \cite{Mangano},
it is known that the experiments will explore 
the central rapidity region (the CMS and ATLAS detectors will cover $y < 0.9$
only) and that bottom can be tagged, but most likely not charm \footnote{Tagging 
is done by finding the point where the quark
decays (called the vertex). The probability of decaying is exponential,
and higher in the region close to the collision point. The only way to tell 
charm from
bottom is by the distance from the collision to the vertex and bottom quark 
lives
longer than charm. Thus, to detect charm with a good degree of confidence, one 
needs to
select vertices close to the collision point.
But in this region the vertices from bottom decay dominate, because the number
 of decay channels of the $b$ quark is five
times larger than that of the $c$ quark.} \cite{GIGI}. 
This means that the lowest $x$ that LHC is expected to reach, assuming
realistically that charm will not be tagged, is 
$x \simeq m_b \exp{(-0.9)}/ \sqrt{s} = 1.5
\times 10^{-4}$.
Therefore, the method proposed here may give information on the gluon PDF at
 $x < 10^{-5}$, a range not reachable in laboratory experiments 
 in the near future. 

To this end we studied the dependence of the leptonic fluxes and their slopes on 
$\lambda$. The slopes depend almost linearly on $\lambda$. We studied the 
uncertainties of the method we propose (excluding the experimental errors of the 
telescopes themselves). These come mainly from two sources: the free parameters 
of the NLO QCD calculation of charm production and the poorly known composition 
of cosmic rays at high energies.
 
We have seen that, for a fixed value of $\lambda$, the uncertainties give rise 
to an error band for the leptonic fluxes of almost one order of magnitude at the 
highest energies. This makes
impossible a determination of $\lambda$ based solely on the absolute values of
the fluxes, therefore we propose using the slopes of the fluxes. In particular 
we are proposing to use {\it down-going muons}, for energies $E_{\mu} \gtrsim 
100~TeV$, where prompt muons dominate over conventional ones, and not up-going 
neutrino-induced muons whose flux is orders of magnitude smaller. 
While an important contribution to up-going muons is expected from astrophysical 
neutrinos, there is no background for down-going atmospheric muons.

The overall theoretical error, from the charm production model, on the 
measurement of $\lambda$, is $(\Delta \lambda)_{charm} \lesssim 0.10$. A 
comparable error, due to uncertainties in the cosmic ray composition, $(\Delta 
\lambda)_{comp} \lesssim 0.15$, must be added, so that the overall error in the 
determination of $\lambda$ with neutrino telescopes is $\Delta \lambda \sim 
0.2$. 

These errors may be reduced by improving the experimental knowledge of the charm 
production cross sections and of the cosmic ray composition
around and above the knee.
 
\bigskip\bigskip 
 
\acknowledgments

The authors would like to thank M. Mangano, P. Nason, L. Rolandi and 
D. Treleani for helpful discussions. We also thank M. Mangano and  P. Nason
for the MNR program. This research was supported in part by the US
Department of Energy under grant DE-FG03-91ER40662 Task C.

\newpage 
 
\section*{FIGURE CAPTIONS} 
 
\begin{description} 

\item[Fig. 1] The ratio
$R = (\sigma_{NLO} - \sigma_{LO}) /
         (\sigma_{LO} \alpha_s \ln(s/m_c^2)/\pi)$ 
is plotted as a function of the beam energy $E$, for the different values of 
$\lambda$ used with the MRST PDF.

\item[Fig. 2] Total cross sections for charm production $\sigma_{c\overline{
      c}}$, up to NLO, calculated with MRST ($\lambda = 0$) and the values of
  $m_c$, $\mu_F$, $\mu_R$ of Table~\ref{table:1}, are compared with recent 
experimental
  values \cite{alves1,fmnr,aguilar88,barlag88,ammar88,kodama91}. For each
  ``band'' in the figures (i.e. for $\mu_{F} = m_{T}/2$, $m_{T}$ and $2m_{T}$) 
the cross sections
  increase with increasing $m_c$ (and correspondingly smaller values of $\mu_R$) 
in Table~\ref{table:1} 
  (Fig. 2b is an enlargement of Fig. 2a).
 
\item[Fig. 3] Results for MRST $\lambda = 0$. The $E_{\mu}^{3}$-weighted
  vertical prompt fluxes, at NLO, are calculated using the values of $m_c$,
  $\mu_F$, $\mu_R$ of Table~\ref{table:1} and compared to the TIG \cite{TIG} 
conventional
  and prompt fluxes. For each ``band'' in the figure (i.e. for $\mu_{F} = 
m_{T}/2$, $m_{T}$ and $2m_{T}$) 
  the fluxes increase with increasing $m_c$ (and correspondingly smaller values 
of $\mu_R$) in Table~\ref{table:1}.
 
\item[Fig. 4] Spectral indices $-b_{\mu}$ of the fluxes plotted in Fig.~3, for
  the MRST $\lambda =0$ case. For each ``band'' in the figure (i.e. for $\mu_{F} 
= m_{T}/2$, $m_{T}$ and $2m_{T}$) 
  the spectral indices decrease with increasing $m_{c}$ (and correspondingly 
smaller values of $\mu_R$) in Table~\ref{table:1}.
 
\item[Fig. 5] Results for MRST $\lambda = \lambda (T)$. The $
  E_{\mu}^{3}$-weighted vertical prompt fluxes, at NLO, are calculated using
  selected values of $m_c$, $\mu_F$, $\mu_R$ from Table~\ref{table:1} and 
compared to the
  TIG \cite{TIG} conventional and prompt fluxes.
 
\item[Fig. 6] Results for MRS R1-R2, CTEQ 4M, MRST, for $\lambda = 0$ (Fig. 6a) 
and $\lambda = 0.5$ (Fig. 6b),
  with standard choice of parameters $m_c$, $\mu_F$, $\mu_R$. Top part: $
  E_{\mu}^{3}$-weighted vertical prompt fluxes, at NLO. Bottom part: related
  spectral indices $-\alpha_{\mu}$ (for the $\lambda = 0$ case, $ -\alpha_{\mu}
  = -b_{\mu}$).
 
\item[Fig. 7] Results for MRST $\lambda =0-0.5$ (solid lines). The $
  E_{\mu}^{3}$-weighted vertical prompt fluxes, at NLO, are compared to the TIG
  \cite{TIG} conventional and prompt fluxes (dashed lines). 

\item[Fig. 8] Results for MRST $\lambda =0-0.5$ (solid lines). The $
  E_{\mu}^{2}$-weighted integrated vertical prompt fluxes, at NLO, are compared
  to the number of particles traversing a km$^{3}~2\pi$ sr detector per year
  (dotted lines). 
  
\item[Fig. 9] Results for MRST $\lambda =0-0.5$. The spectral
  indices $-\alpha _{\ell }(E_{\ell }) $ for the different values of $\lambda
  $, calculated directly by our simulation (solid lines) are compared to the
  corresponding terms $-b_{\ell }(E_{\ell })+\lambda $ (dotted lines). 
  
\item[Fig. 10] Results for MRST $\lambda =0-0.5$ (solid lines). The error of
  Eq.~(\ref{eq:4.24}) is evaluated in terms of the difference $\alpha _{\ell
    }(E_{\ell })-b_{\ell }(E_{\ell })+\lambda $.

\item[Fig. 11] Results for MRST $\lambda =0$ for different values of $\gamma$. 
a) The spectral indices $-\alpha_{\ell}(E_{\ell})$ for the different values of 
$\gamma$, calculated directly by our simulation (solid lines) are compared to 
the corresponding terms $-{\bar b}_{\ell}(E_{\ell}; \gamma=1.7,~2.0; 
\lambda=0)-\gamma$, with increments in $\gamma$ equal to $\pm0.1,~\pm0.2$ 
(dotted lines). The curves are labelled by the related value of $\gamma$ above 
the knee ($\gamma=2.0$ is our ``standard value''). b) Uncertainty due to the 
non-linearity of Eq.~(\ref{eq:4.30}), as the difference $\alpha_{\ell}-{\bar 
b}_{\ell}-\gamma$.

\item[Fig. 12] The $E_{N}^{3}$-weighted equivalent nucleon flux $\phi_{eq}(E_N)$ 
is shown for different primary cosmic ray experiments 
\cite{JACEE,CASA-MIA,HEGRA,BIERMANN}. For each of these we plot the central 
value and the related 
error band.

\item[Fig. 13] The spectral index, $\gamma_{eq}+1$, for the equivalent nucleon 
fluxes of Fig.~12, is shown for different primary cosmic ray experiments 
\cite{JACEE,CASA-MIA,HEGRA,BIERMANN}. For each of these we plot the central 
value and the related error band.

\end{description} 
 
\newpage 
 
\begin{table}[tbp] 
\begin{center} 
\begin{tabular}{|l|c|c|c|c|c|} 
\hline 
&  &  &  &  &  \\  
& $m_c~({\rm GeV})$ & $\mu_{F}~(m_{T})$ & $\mu_{R}~(m_{T})$ & $\sigma_{c \bar 
c}^{MNR}~(\mu {\rm b})$ & $\sigma_{c\bar c}^{EXP}~(\mu {\rm b})$ \\  
&  &  &  &  &  \\ \hline 
& $1.1$ & $0.5$ & $2.53$ & $13.48$ & $13.5 \pm2.2$ \\  
&  & $1.0$ & $2.40$ & $13.48$ & " \\  
&  & $2.0$ & $2.10$ & $13.42$ & " \\ \hline 
& $1.2$ & $0.5$ & $1.46$ & $13.57$ & $13.5 \pm2.2$ \\  
&  & $1.0$ & $1.40$ & $13.54$ & " \\  
&  & $2.0$ & $1.23$ & $13.51$ & " \\ \hline 
& $1.25$ & $0.5$ & $1.18$ & $13.57$ & $13.5 \pm2.2$ \\  
&  & $1.0$ & $1.13$ & $13.54$ & " \\  
&  & $2.0$ & $1.00$ & $13.58$ & " \\ \hline 
& $1.3$ & $0.5$ & $0.96$ & $13.55$ & $13.5 \pm2.2$ \\  
&  & $1.0$ & $0.92$ & $13.50$ & " \\  
&  & $2.0$ & $0.83$ & $13.53$ & " \\ \hline 
& $1.4$ & $0.5$ & $0.68$ & $13.51$ & $13.5 \pm2.2$ \\  
&  & $1.0$ & $0.66$ & $13.51$ & " \\  
&  & $2.0$ & $0.61$ & $13.52$ & " \\ \hline 
\end{tabular} 
\end{center} 
\caption{ 
Choice of parameters $m_{c}$, $\mu_{F}$ and $\mu_{R}$, that can reproduce 
the experimental total cross section $\sigma _{c\bar{c}}^{EXP}$ for charm 
production in $pN$ collisions at 250 GeV from the E769 experiment. For each 
set of parameters, $\sigma _{c\bar{c}}^{MNR}$ is the cross section 
calculated with the MNR program using MRST PDF.} 
\label{table:1} 
\end{table} 

\newpage
\begin{figure}[t]
\epsfig{file=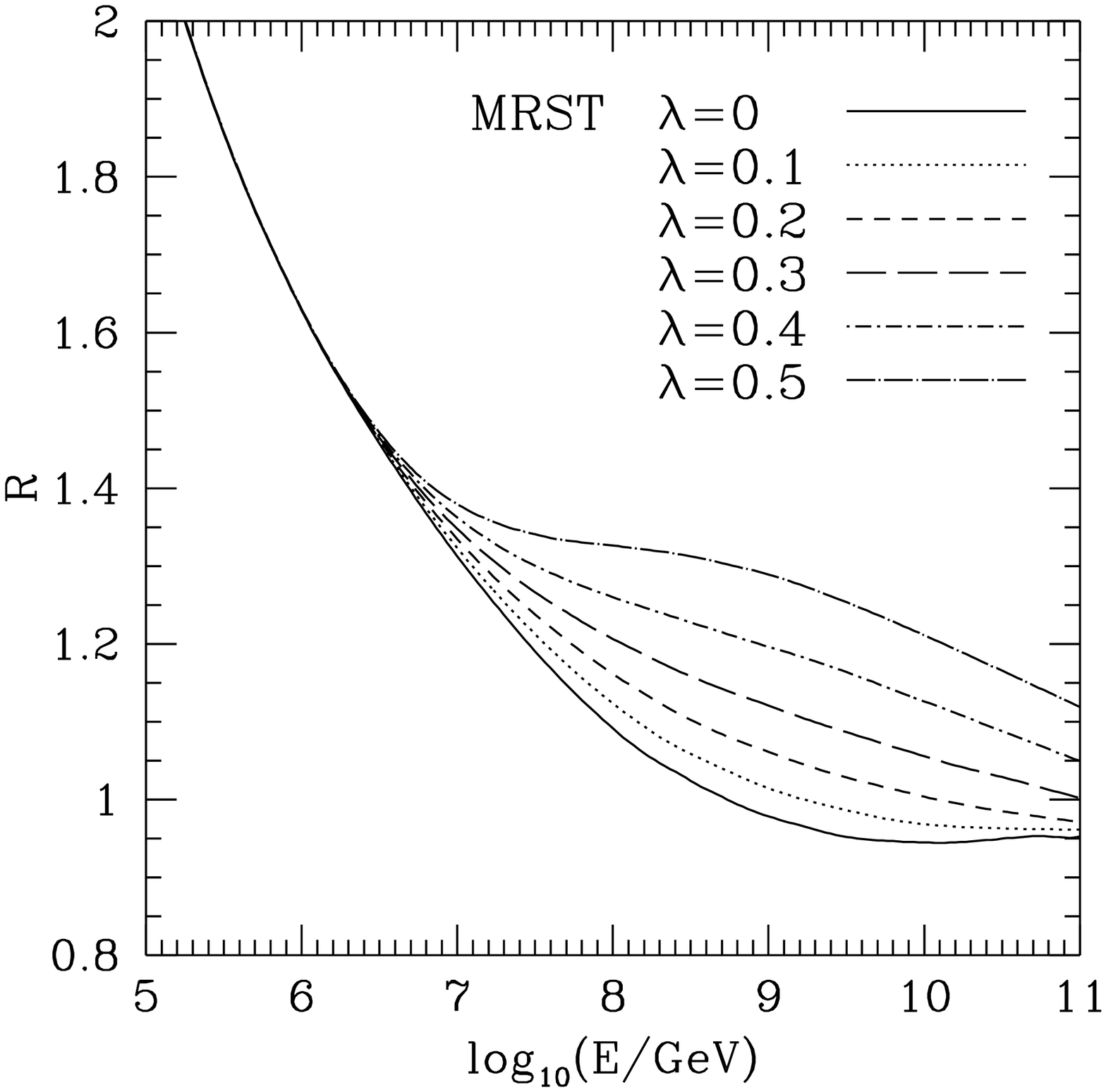,width=\textwidth}
Figure 1.
\end{figure}

\newpage
\begin{figure}[t]
\epsfig{file=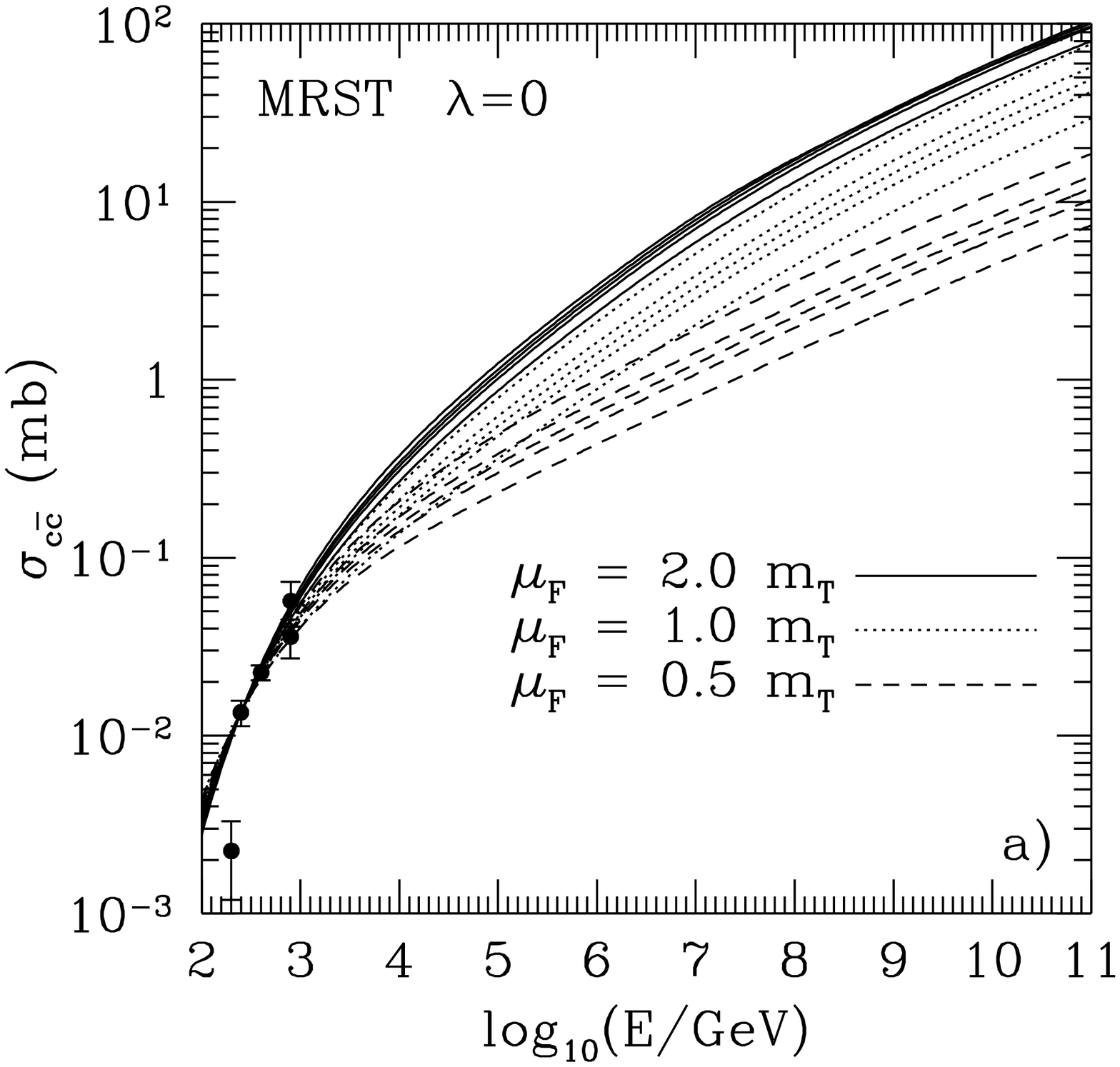,width=\textwidth}
Figure 2a.
\end{figure}

\newpage
\begin{figure}[t]
\epsfig{file=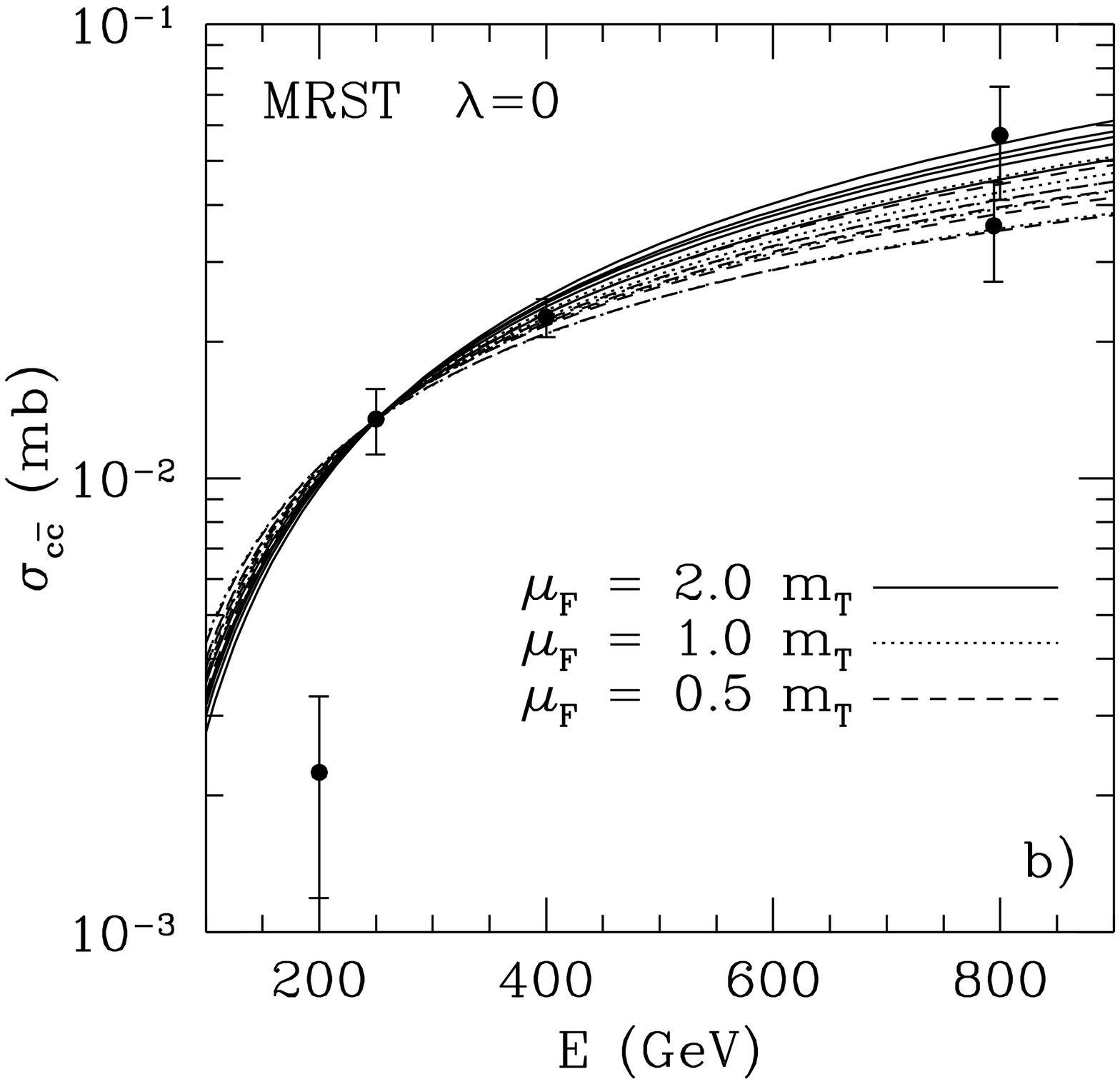,width=\textwidth}
Figure 2b.
\end{figure}

\newpage
\begin{figure}[t]
\epsfig{file=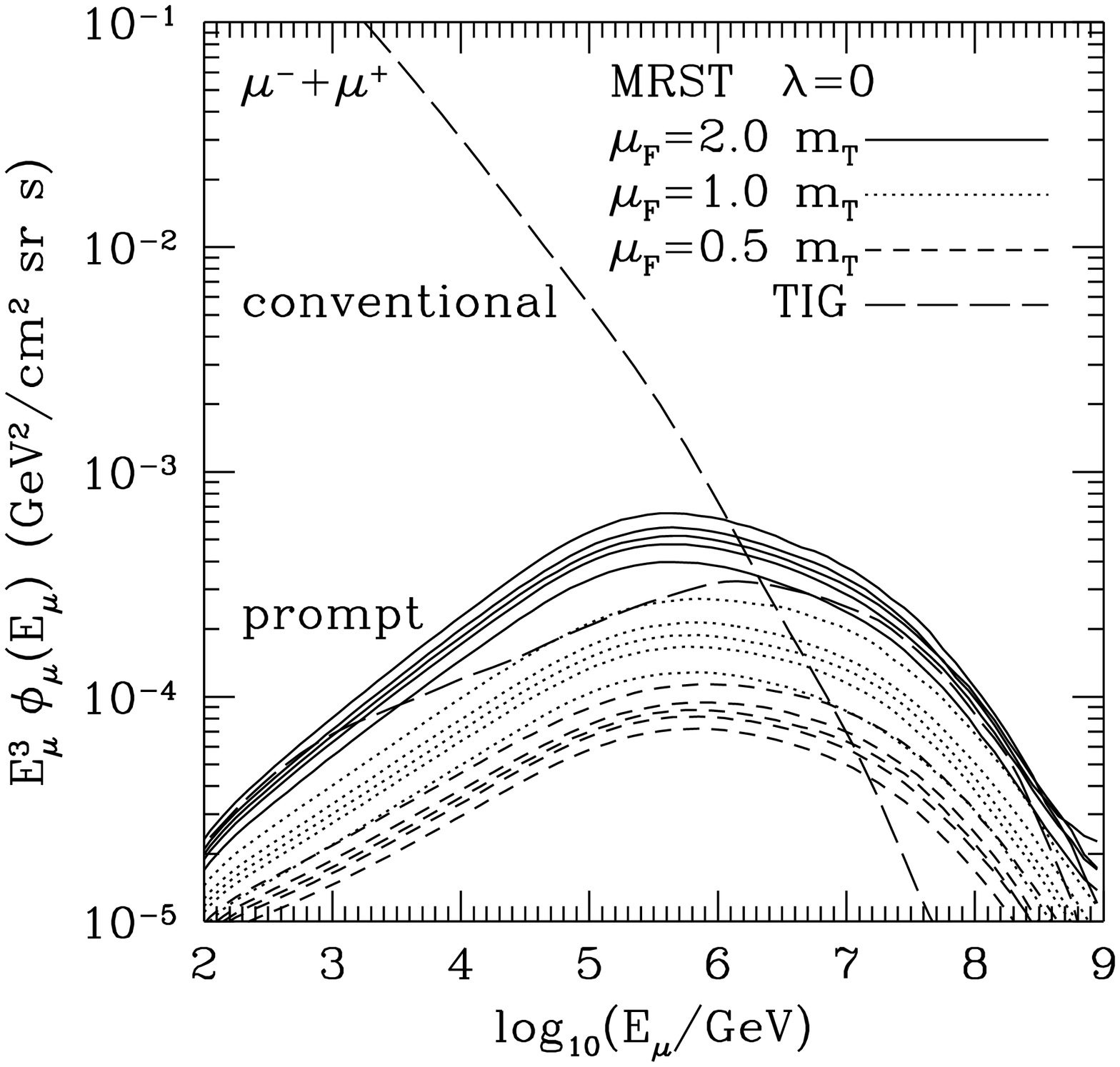,width=\textwidth}
Figure 3.
\end{figure}

\newpage
\begin{figure}[t]
\epsfig{file=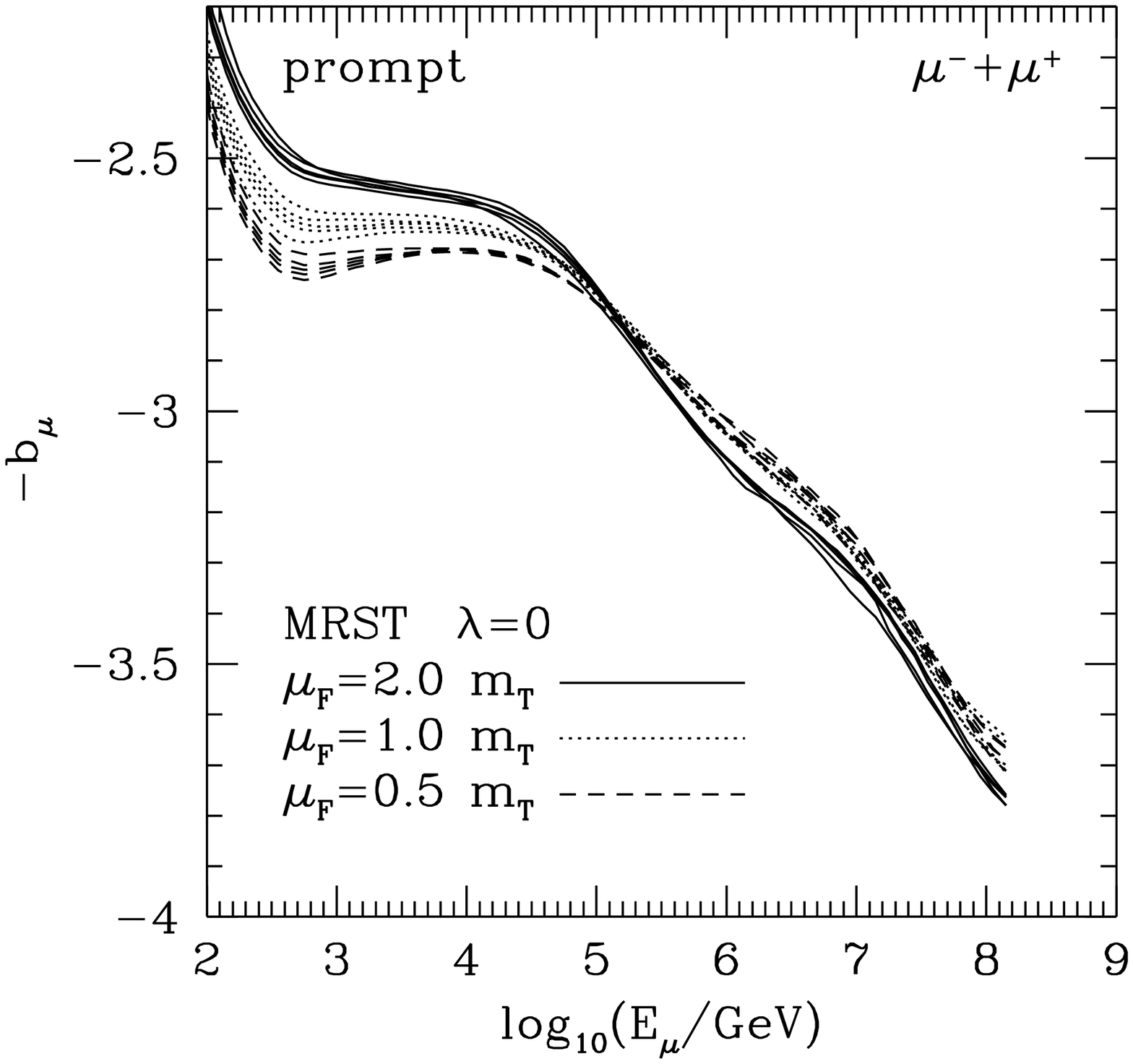,width=\textwidth}
Figure 4.
\end{figure}

\newpage
\begin{figure}[t]
\epsfig{file=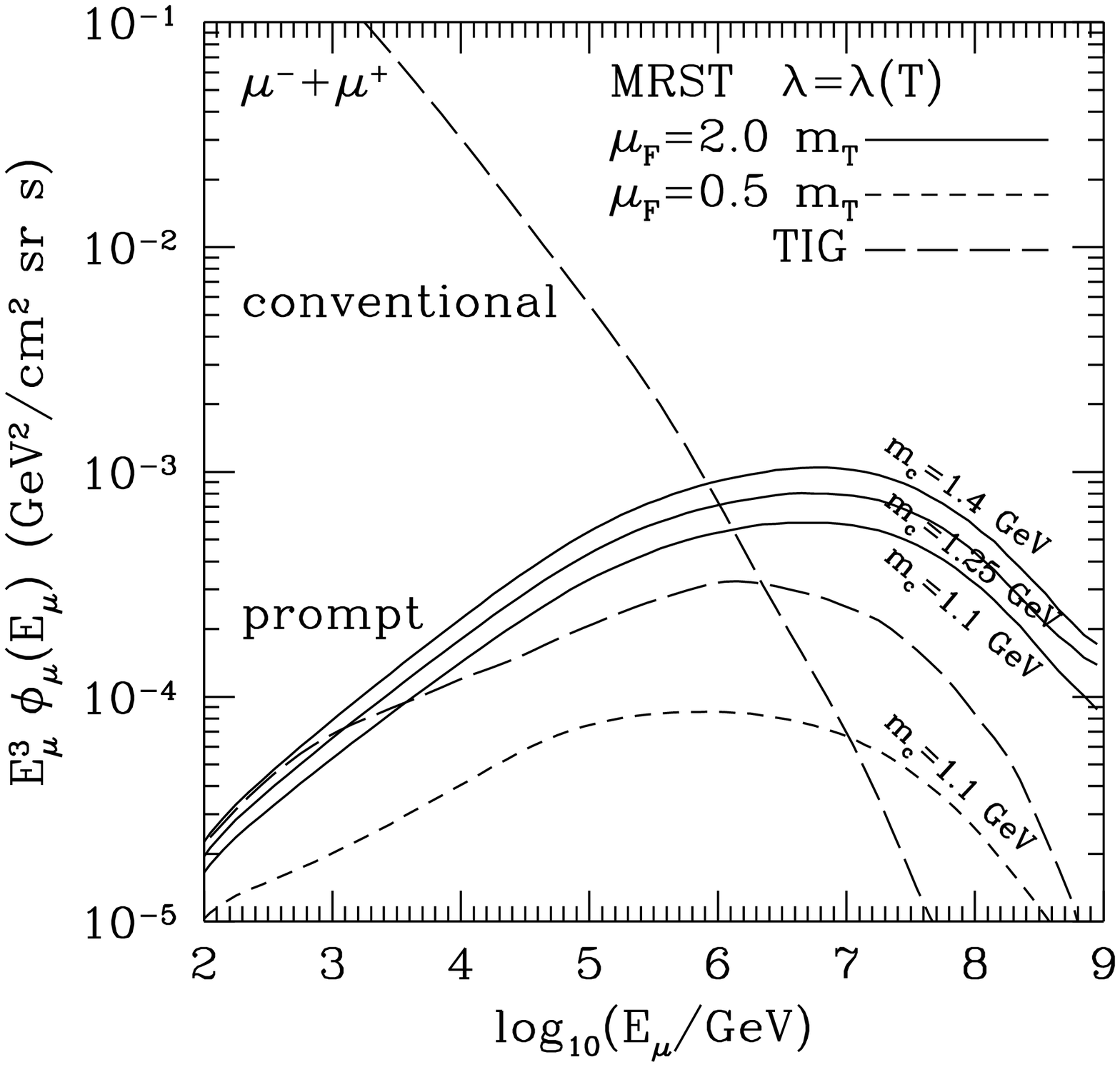,width=\textwidth}
Figure 5.
\end{figure}

\newpage
\begin{figure}[t]
\epsfig{file=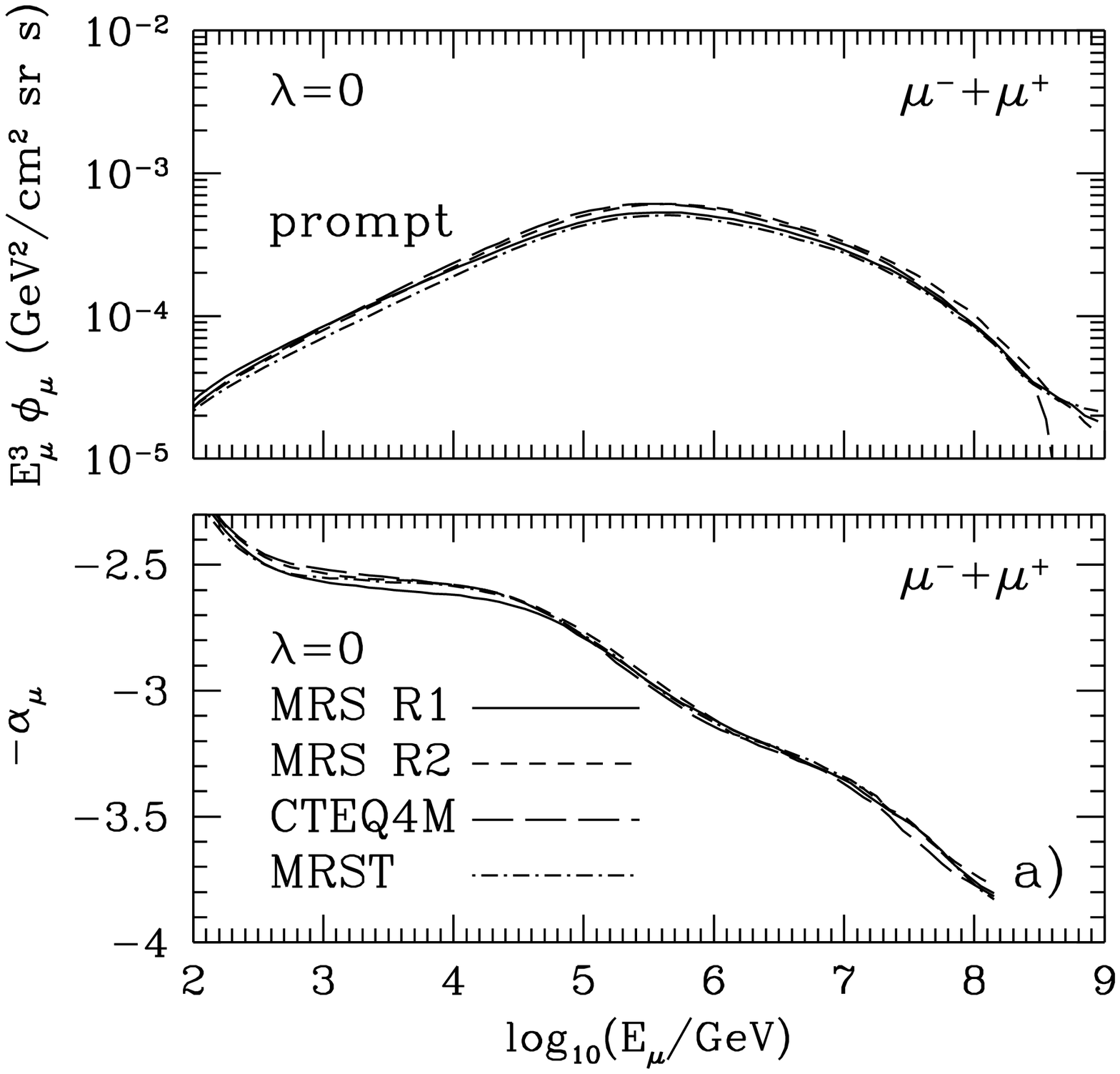,width=\textwidth}
Figure 6a.
\end{figure}

\newpage
\begin{figure}[t]
\epsfig{file=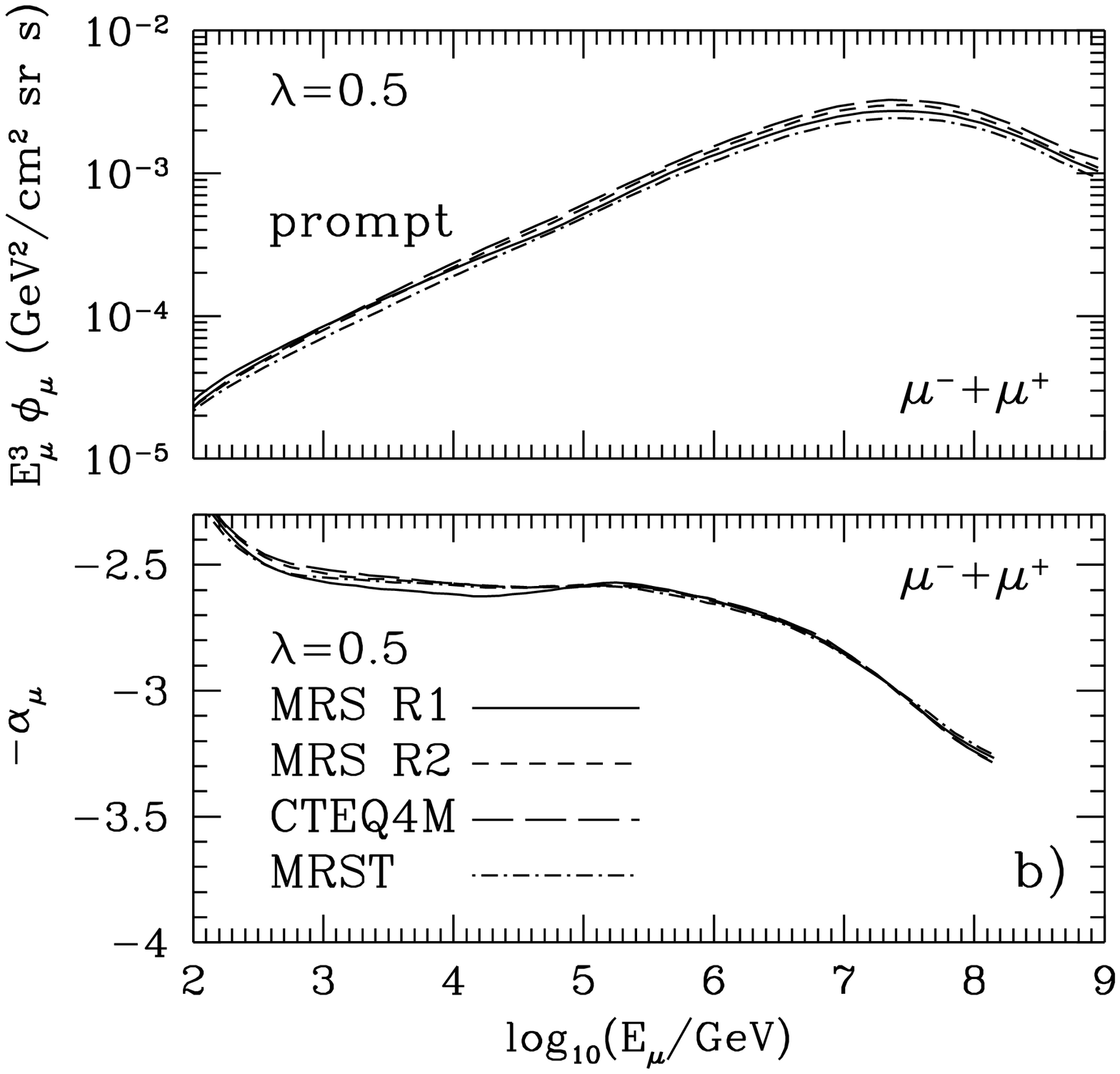,width=\textwidth}
Figure 6b.
\end{figure}

\newpage
\begin{figure}[t]
\epsfig{file=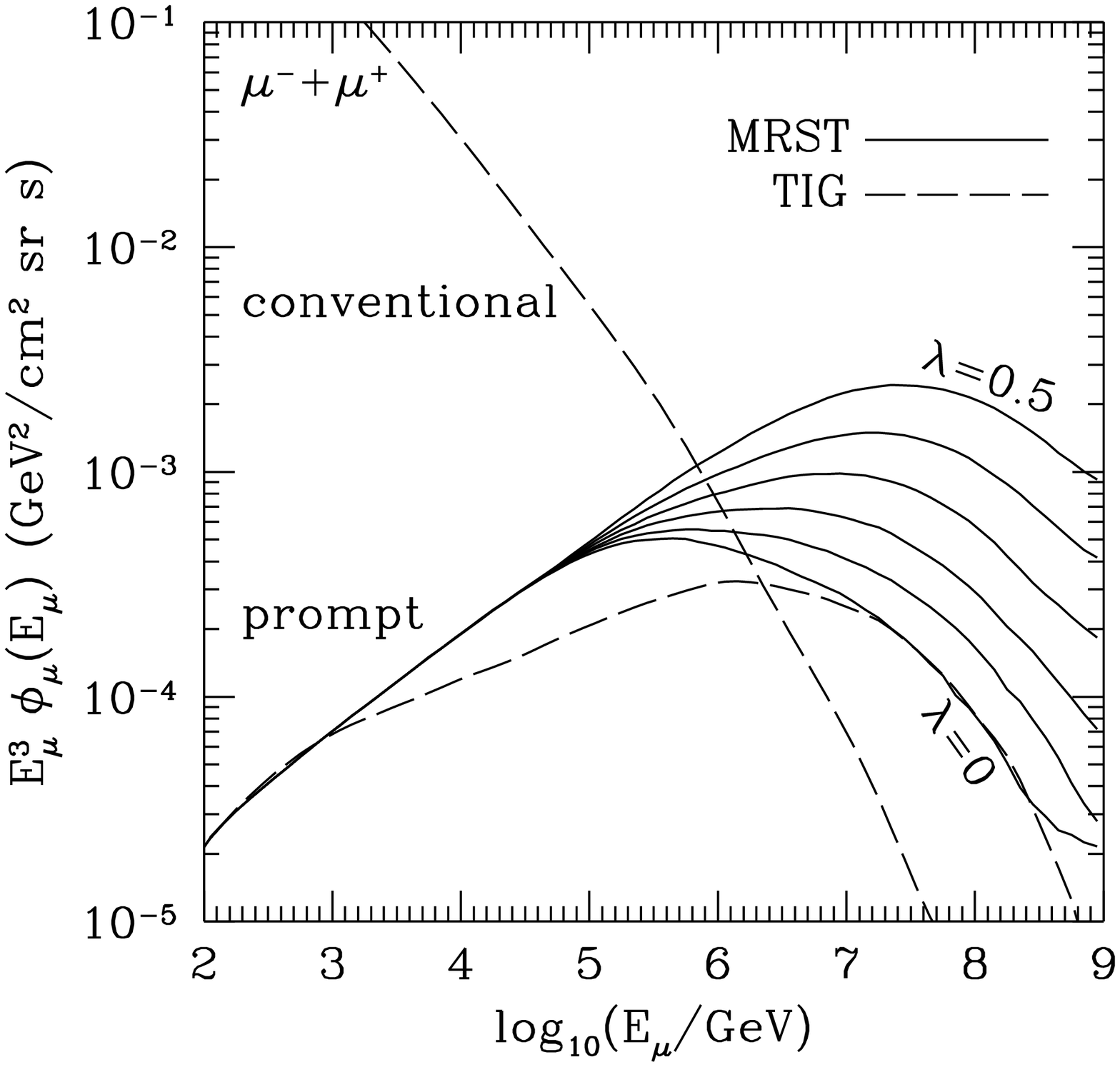,width=\textwidth}
Figure 7.
\end{figure}

\newpage
\begin{figure}[t]
\epsfig{file=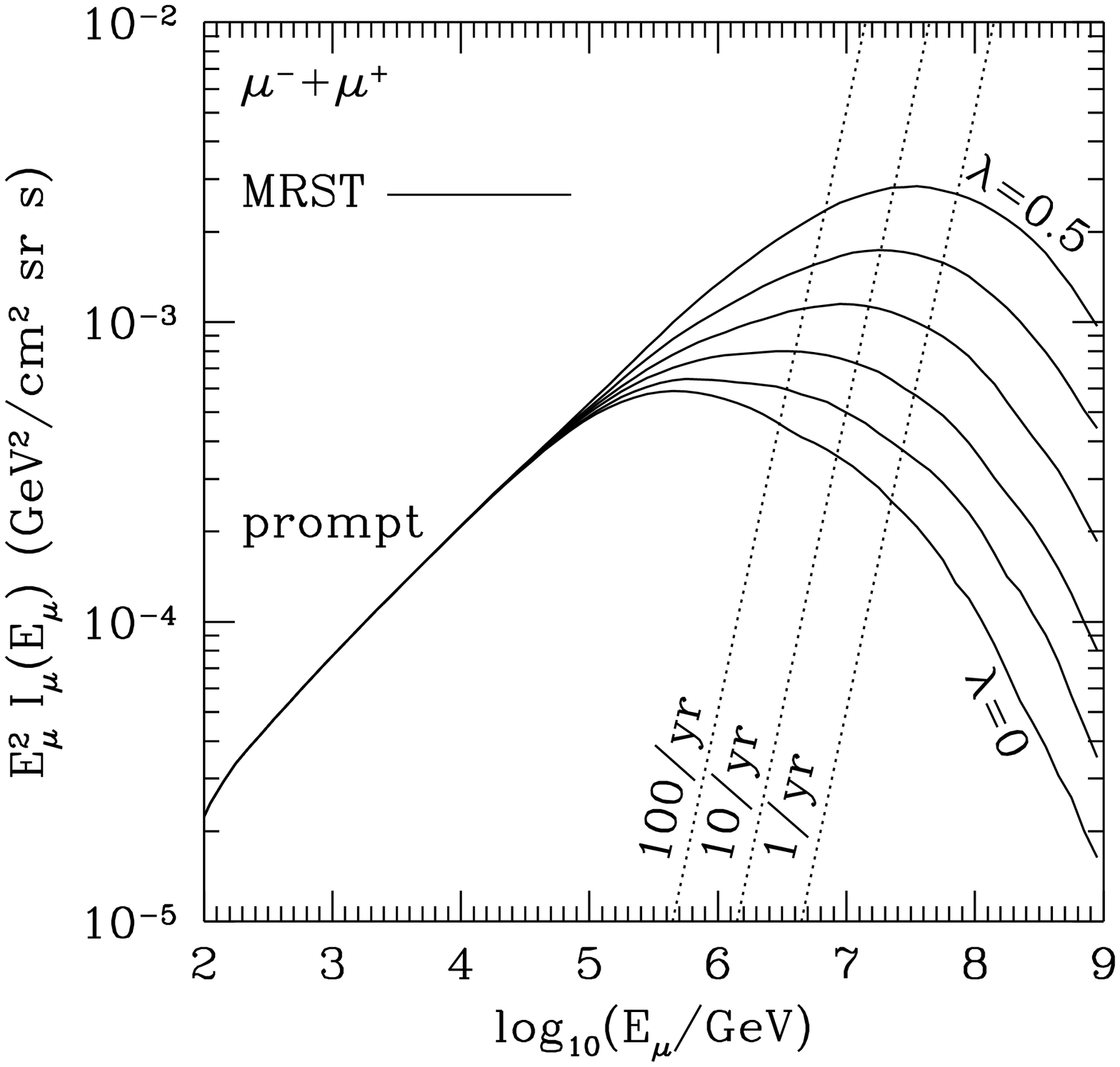,width=\textwidth}
Figure 8.
\end{figure}

\newpage
\begin{figure}[t]
\epsfig{file=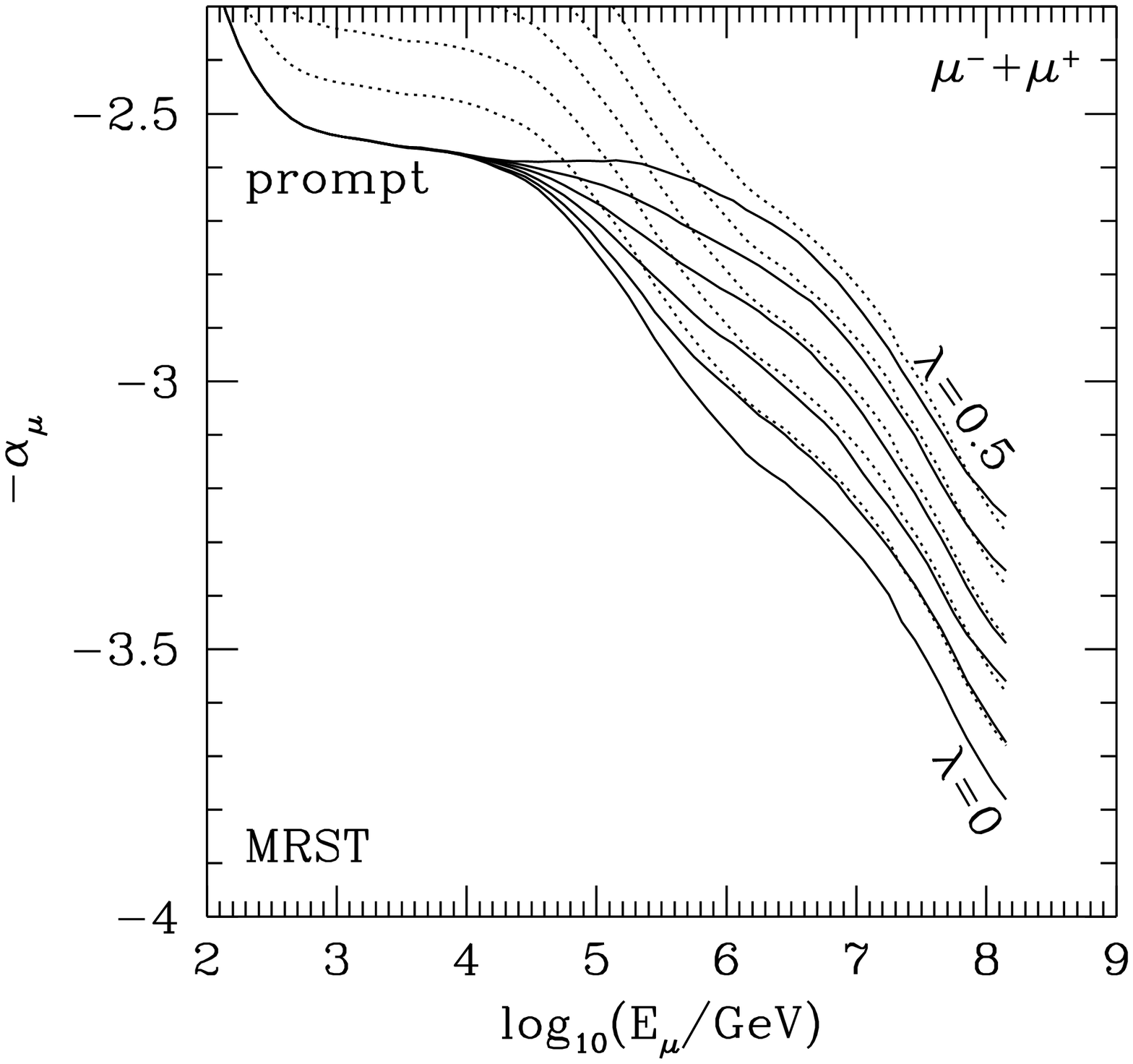,width=\textwidth}
Figure 9.
\end{figure}

\newpage
\begin{figure}[t]
\epsfig{file=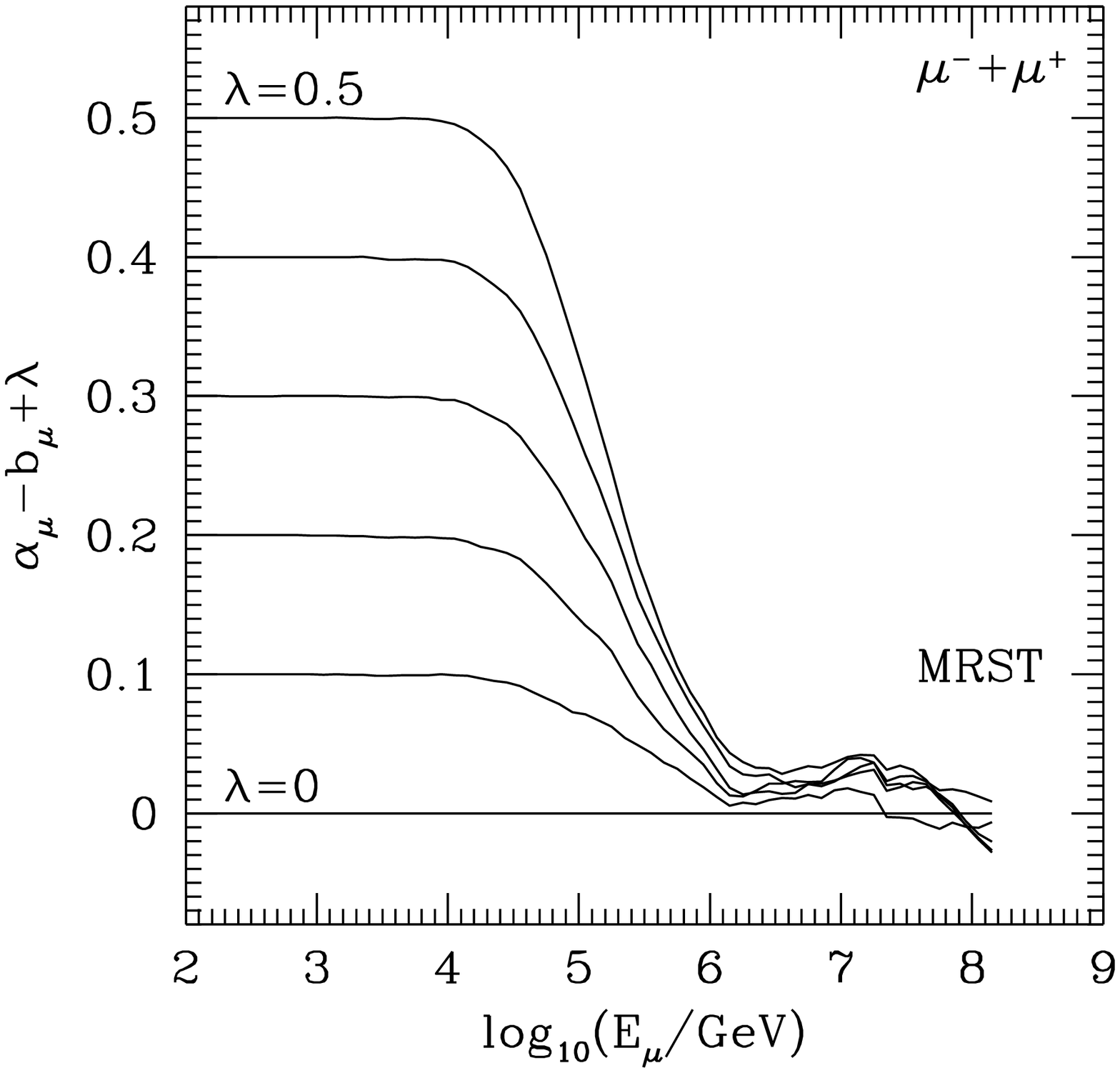,width=\textwidth}
Figure 10.
\end{figure}

\newpage
\begin{figure}[t]
\epsfig{file=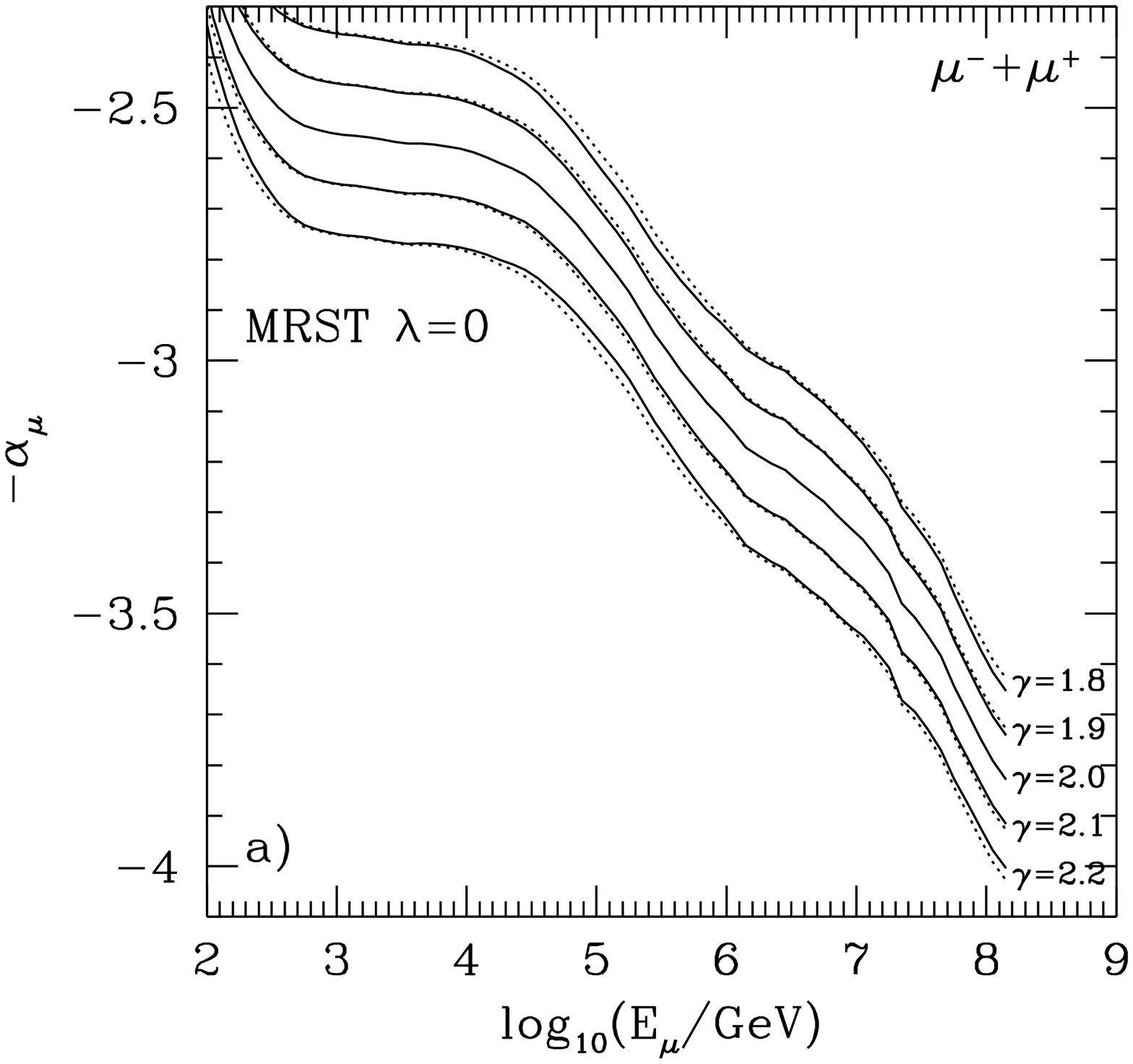,width=\textwidth}
Figure 11a.
\end{figure}

\newpage
\begin{figure}[t]
\epsfig{file=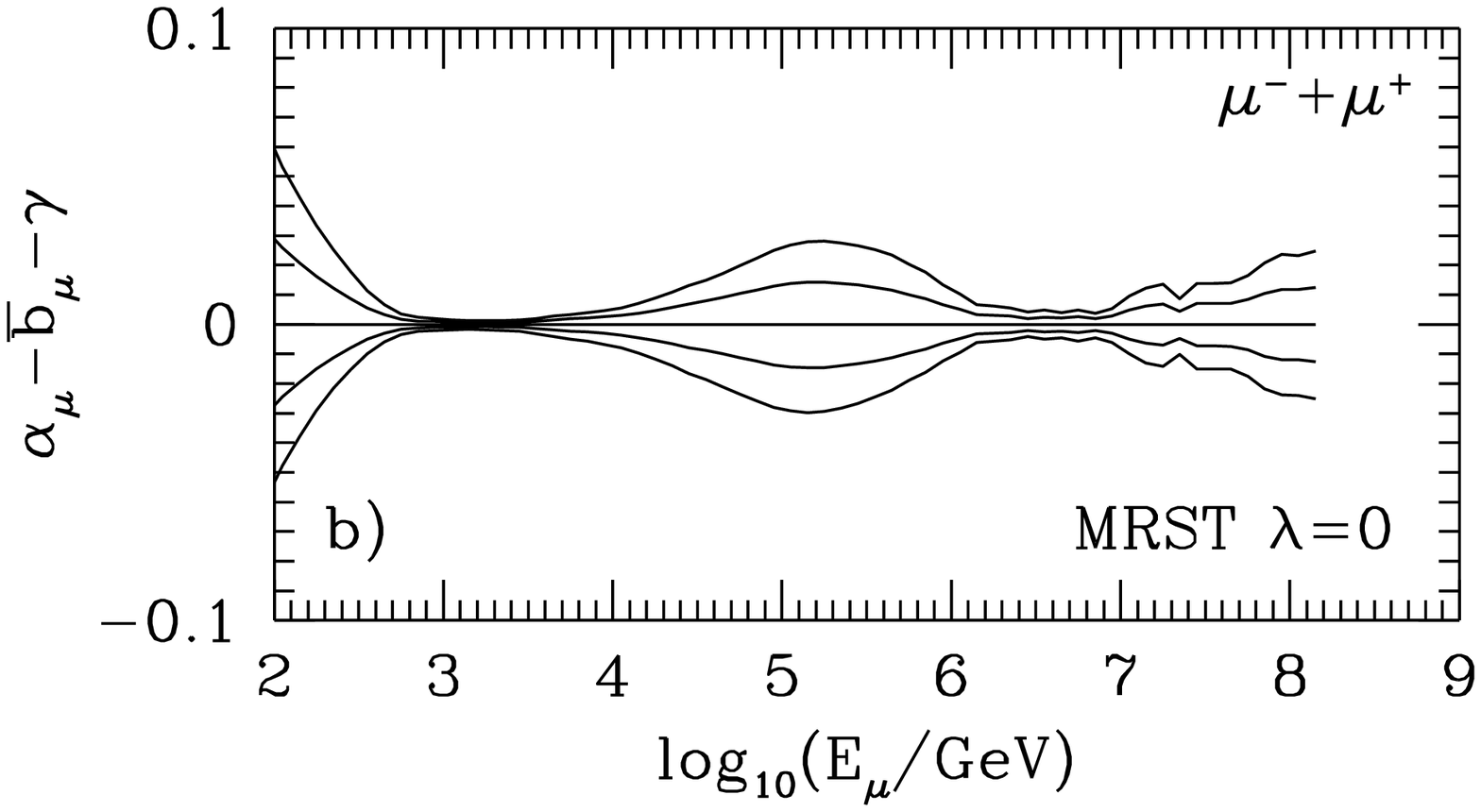,width=\textwidth}
Figure 11b.
\end{figure}

\newpage
\begin{figure}[t]
\epsfig{file=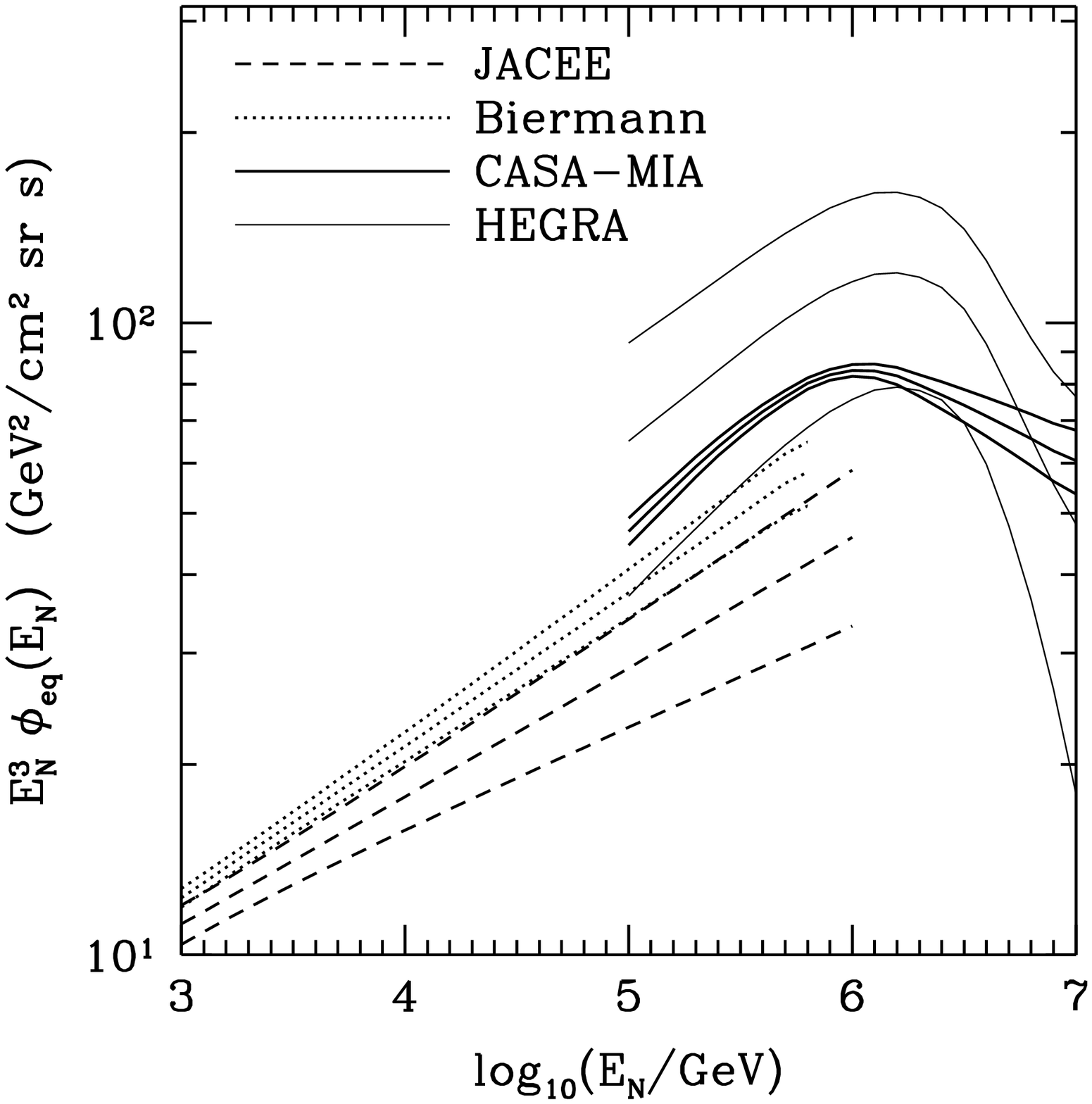,width=\textwidth}
Figure 12.
\end{figure}

\newpage
\begin{figure}[t]
\epsfig{file=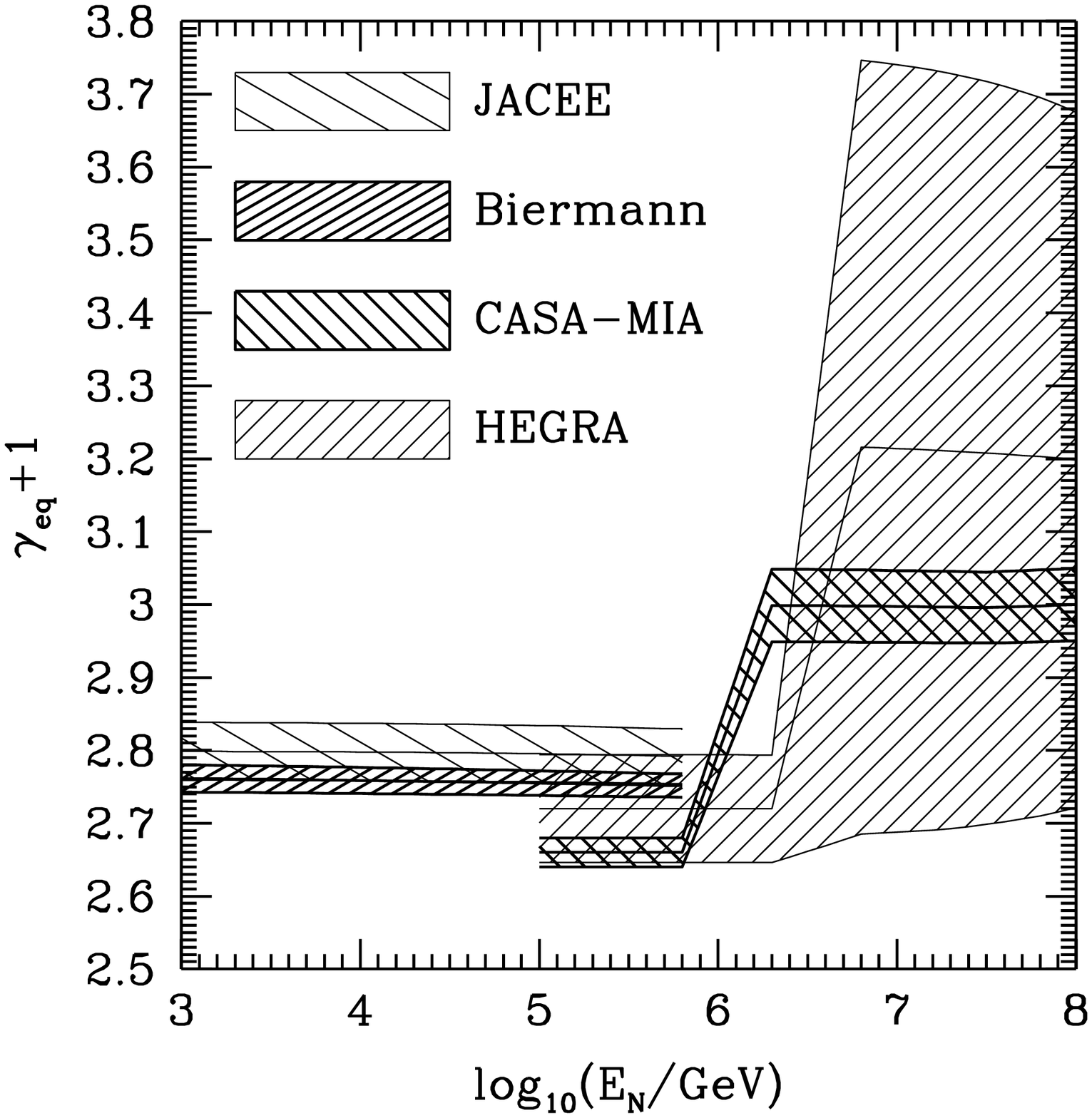,width=\textwidth}
Figure 13.
\end{figure}

\end{document}